\documentclass[12pt]{article}
\usepackage{epsf}
\def\G{{\cal G}^{+++}}
\begin{document}
\thispagestyle{empty}
\setcounter{page}{0}
\renewcommand{\theequation}{\thesection.\arabic{equation}}

{\hfill{\tt hep-th/0311255}}

{\hfill{ULB-TH/03-38}}

\vspace{1.5cm}

\begin{center} {\bf    $\G$ INVARIANT FORMULATION OF GRAVITY AND    M-THEORIES:

EXACT BPS SOLUTIONS}

\vspace{.5cm}

Fran\c cois Englert${}^a$ and Laurent Houart${}^b$\footnote{Research   Associate
F.N.R.S.}

\footnotesize \vspace{.5 cm}

${}^a${\em Service de Physique Th\'eorique\\ Universit\'e Libre de   Bruxelles,
  Campus Plaine, C.P.225\\Boulevard du Triomphe, B-1050 Bruxelles,   Belgium\\
and\\ Racah Institute of Physics\\ Hebrew University of Jerusalem, 91904  
Jerusalem, Israel}\\ {\tt fenglert@ulb.ac.be}

\vspace{.2cm}

${}^b${\em Service de Physique Th\'eorique et Math\'ematique }\\  {\em
Universit\'e Libre de Bruxelles, Campus Plaine C.P. 231}\\ {\em   Boulevard du
Triomphe, B-1050 Bruxelles, Belgium}\\ {\tt lhouart@ulb.ac.be}

\end{center}

\vspace {1cm}
\centerline{ABSTRACT}
\vspace{- 3mm}
\begin{quote}\small We present a tentative formulation of  theories of
gravity with suitable matter content, including in particular  pure gravity in
$D$ dimensions,   the bosonic effective actions of M-theory and  of   the
bosonic string,  in terms of actions invariant under very-extended Kac-Moody  
algebras
$\G$.  We conjecture that they host additional degrees of freedom not  
contained in the conventional theories. The actions are  constructed  in a  
recursive way from a level expansion for all very-extended algebras $\G$.
They   constitute non-linear realisations on  cosets, a priori unrelated to
space-time,   obtained from a  modified Chevalley involution.   Exact solutions
are found for all
$\G$. They describe the algebraic  properties of BPS extremal branes,  
Kaluza-Klein waves and Kaluza-Klein monopoles. They illustrate the
generalisation to   all
$\G$ invariant theories of the well-known duality properties of string  
theories by expressing duality  as Weyl invariance in $\G$. Space-time is
expected   to be generated dynamically. In the level decomposition  of
$E_8^{+++}=E_{11}$, one may indeed select an  $A_{10}$  representation  of 
generators $P_a$  which appears  to engender space-time
translations by inducing infinite towers of fields interpretable as field
derivatives in space and time.

\end{quote}

\newpage
\baselineskip18pt

\setcounter{equation}{0}
\addtocounter{footnote}{-1}
\section{Introduction}

String theories, and  particularly superstrings and their possible  
unification at the non-perturbative level in an elusive M-theory, are often
viewed in the   double perspective of  a consistent gravity theory  and   of
fundamental interaction unification. Matter degrees of freedom relevant   for
the unification program  are identified with the huge number of degrees of
freedom present in the string modes. Consistency with quantum gravity
requirements at the perturbative level and   impressive   theoretical
successes, among which the  evaluation of the statistical entropy of    near
extremal black holes \cite{mald} is probably the
most significant, seem to indicate    that the string-M-theory approach
contains elements of a consistent theory of   gravity and matter.

The  project however stumbles when confronting gravity at the non perturbative
level. Our point of view is that, although the   introduction  of the new
degrees of freedom in string theories  is essential and constitutes   a clue of
a correct computation of the black hole entropy, some new physical ingredient, 
which is to some extend foreign to the unification paradigm,  is needed to cope
with gravity.  The present   work is an attempt to uncover such ingredient.

We take advantage of two different trends.

The first one is the interpretation of the well-known symmetry of the scalar
cosets emerging in the dimensional reduction of gravity,   suitably coupled to
forms and to dilatons,  as a remnant of a much larger symmetry of the full
covariant and gauge invariant original theory. Coset symmetries were first
found in  the dimensional reduction of eleven-dimensional supergravity
\cite{cremmerjs78} but appeared also in other theories. They have been the
subject of much study, and some classic example are given in \cite{allcoset}.
In fact, all
  simple maximally non-compact Lie group $\cal G$   could be generated from the
reduction down to three dimensions of suitably chosen actions
\cite{cremmerjlp99}. It was conjectured that these actions, or some unknown
extension of them, possess the much   larger very-extended Kac-Moody symmetries
$\G$. $\G$ algebras are defined from the Dynkin diagrams obtained from   those
of
$\cal G$ by adding three nodes \cite{olivegw02}.
  One first adds the affine node, then a second node connected  to it by a
single line and then similarly a third one   connected to the second. These
define respectively the affine Kac-Moody algebra
$\cal G^+$,  the overextended and very extended  Lorentzian Kac-Moody algebras
$\cal G^{++}$ and $\G$. The conjecture of a $\G$ symmetry originated from the
study of    generalisations to gravity and to form field strengths of  the
considerations which    produced the scalar cosets. The $E_8$ invariance of the
dimensional reduction to   three dimensions of 11-dimensional supergravity
would be enlarged to  
$E_8^{+++} =E_{11}$, as first proposed in reference \cite{west01}. Similarly the
effective action of the bosonic string  would have the symmetry
$D_{24}^{+++}=k_{27}$ \cite{west01}. $A_{D}^{+++}$ was also proposed  
\cite{lambertw01} for pure gravity in $D$ space-time dimensions. It was then
shown  that some  solutions of general relativity and dilatons form
representations of the Weyl group   of $\G$ for {\em all} actions $S$, which
dimensionally reduced to three dimensions   lead to a Lie group $\cal G$
symmetry, and the extension to  $\G$ was proposed in   general \cite{ehtw}. In
a different development, the study of the  properties of   cosmological
solutions in the vicinity of a space-like singularity, known as cosmological  
billiards 
\cite{damourhn00}, revealed an overextended symmetry
$\cal G^{++}$  for all $\cal G$  
\cite{{damourh00},damourbhs02}.

The second one stems from  the remarkable attempt to recover,
in the case of M-theory, the bosonic equations of motion of supergravity out of
the overextended   symmetry
$E_8^{++}=E_{10}$ \footnote{In the context of dimensional reduction,   the
appearance of
$E_{10}$ in one dimension has been first conjectured by B. Julia  
\cite{julia84}.}. An
$E_8^{++}$-invariant  Lagrangian was proposed \cite{damourhn02}. It was built
in a recursive way on the coset $E_8^{++}/K^{++}$, where
$K^{++}$ is the subalgebra of
$E_8^{++}$ invariant under the Chevalley involution, by a `level'   expansion
in terms of the subalgebra $A_9$. The level of an irreducible representation
of  
$A_9$ counts the number of times the special root $\alpha_{11}$ in the Dynkin  
diagram of
$E_8^{++}$ (that is  the $E_8^{+++}$ diagram of Fig.1 with the node 1 erased)
appears in the decomposition of the adjoint representation of  
$E_8^{++}$ into
$A_9$ representations. The theory was formulated as a perturbation   expansion
  near a space-like singularity and checked  up to the third level. The
supergravity fields  were taken to depend
  on time only. Space derivatives  were supposed to be hidden in higher level 
objects, together with new degrees of freedom hopefully related to the Hilbert
space of   superstrings.

The overextended $\cal G^{++}$ leaves out naturally  time as the special
coordinate, which was explicitly introduced in the non-linear   realisation of
$E_8^{++}$. Here, we will formulate a  $\G$ invariant theory by putting all
space-time coordinates on the same footing by defining a   map to $\G$ of a
world-line  {\em a priori unrelated  to space-time}. The   latter should then be
deduced dynamically.  Such an approach to gravity and forms, if successful,
  would  dispose of the need of   explicit
diffeomorphism invariance or  gauge invariance.  All such
information should be hidden in the {\em global}
$\G$ invariance. Although it may seem that global symmetries cannot   contain
local symmetries, in particular in view of the celebrated Elitzur theorem
\cite{elitzur75}, this need not be the case in view of the infinite number of
generators of  
$\G$. We formulate the $\G$ invariant theory from a level  
decomposition\footnote{Level expansions of very-extended algebras in terms of
the subalgebra $A_{D-1}$ have been considered in \cite{west02,west05}.}
  with respect to a subalgebra $A_{D-1}$ where $D$  turns out to be   the
space-time dimension. Our formulation is exploratory and does not pretend to be
a final one.  No attempt is made to cope with   fermionic degrees of
freedom and we limit here our investigation  to the classical   domain. Note
however that the very fact that the theory is not formulated in   space time
and that it rests on a huge symmetry  opens new perspectives for the  
quantisation procedure.

To test the validity of our approach, we derive  and discuss exact `BPS'
solutions of the $\G$  invariant action   and compare them to BPS solutions of
 maximally oxidised theories  discussed in
\cite{ehw}. We obtain in this way the full algebraic structure of the BPS
solutions   of these conventional actions and put into evidence their
group-theoretical significance.   This approach does not yield direct
information about  their space-time  behaviour   but we find indications on 
how space-time  can be encoded in the  
$\G$ invariant theories. One can indeed select in the level decomposition of
$E_8^{+++}=E_{11}$ an $A_{10}$  representation of
 generators $P_a$ that appears to engender space-time
translations. A more detailed analysis of a dictionary
translating the content of $\G$   into space-time language is differed to a
separate publication \cite{ehtp}.  Also the   relation between our approach and
the  `hamiltonian' overextended $\cal G^{++}$-theories, in   which time is
explicitly introduced,  will be discussed elsewhere \cite{ehhp}.

The paper is organised as follows. In Section 2, we construct in a recursive
way, for any $\G$, an action invariant under non-linear transformations of
$\G$. The action is defined in a reparametrisation invariant way  on a  
world-line, a priori unrelated to space-time, in terms of fields
$\phi(\xi)$ where $\xi$ spans the world-line.  We
  use a level decomposition of
$\G$ with respect to its subalgebra $A_{D-1}$ where the integer $D$ is  
related to
  the rank  of $\G$. The fields $\phi(\xi)$ live in a coset space
$\G/K^{+++}$ where the subalgebra $K^{+++}$ is invariant under a   `temporal
involution' which is different from the often used Chevalley   involution. The
temporal involution preserves  the Lorentz algebra
$SO(D-1,1)$. As a consequence, the action is Lorentz invariant at each level 
and all fields, which can be transformed between themselves by general
$\G$ transformations   live on a coset
$GL(D)/SO(D-1,1)$. This allows the identification of each $\xi$-field to a
field defined at a {\em fixed} space-time point, independently of  
$\xi$. In Section 3, we find exact solutions of our invariant action for all  
$\G$. These completely define the algebraic properties of extremal BPS branes,  
Kaluza-Klein waves and Kaluza-Klein monopoles (Taub-NUT space-times) \cite{ehw}
and   the motion in
$\xi$-space is consistent with a motion in  the space of   solutions. 
The transformations under the $\G$ Weyl group of these solutions put into  
light the generality of duality transformations which are often considered as  
characteristic of string theories and supersymmetry  
\cite{elitzurgkr97,obersp98,banksfm98}. The only element missing is the
space-time dependance of a harmonic function entering the space-time
description of the solutions. However, we find  indications that  
space-time properties are included in higher levels. 
An $A_{10}$  representation,   
 possibly related to
space-time translations  by   inducing infinite towers of fields interpretable
as field derivatives in space and time, is  exhibited for
$E_8^{+++}=E_{11}$ at level seven. In the concluding section 4, we summarise
and   discuss our results. Detailed computations, in particular of Weyl
reflections in all $\G$, and  illustrations  of the step operator
transformations as duality transformations in $E_8^{+++}$  are given in the
Appendix.

\setcounter{equation}{0}
\section{Non-linear realisation of very extended algebras}

\subsection{Preliminaries}

Theories of gravity in $D$ space-time dimensions coupled to $q$ scalar  
`dilatons' and
$p_I$-form field strengths exhibit, upon dimensional reduction to three  
space-time dimensions, the global symmetry
$GL(D-3)\times U(1)^q$ of deformations of the compact $(D-3)$ torus and   of
dilaton field translations.  For  well chosen
$D, p_I,q$ and dilaton couplings to the
$p_I$-forms, the group $GL(D-3)\times U(1)^q$ of rank $(D-3 +q)$ in   enhanced
to a simple maximally non-compact Lie group $\cal G$ of the same rank. The  
scalar fields of the reduced Lagrangian form a non-linear realisation of $\cal  
G$. They live on the coset space\footnote{For a
general   pedagogical review see \cite{pope}.} $\cal G/H$ where
$\cal H$ is the maximal compact subgroup of   $\cal G$.

For each group $\cal G$, we select a maximally oxidised theory\footnote{We
consider here only theories expressible in Lagrangian form.}, that is   a theory
that is not the dimensional reduction of a higher dimensional one and   has the
symmetry 
$\cal G$ when reduced to three dimensions
\cite{cremmerjlp99}. We write the corresponding Lagrangian as
\begin{equation}
\label{oxid}
  S= \frac{1}{16\pi G_N^{(D)}} \int d^D x \sqrt{-g}
\left[ R - \frac{1}{2} \sum_{u=1}^q (\partial \phi^u)^2 -
\frac{1}{2}\sum_I \frac{1}{ p_I !}\exp(\sum_{u=1}^q a_I^u  
\phi^u)F_{p_I}^2 \right] + C.S.\, ,
\end{equation} where $C.S.$ represents   Chern-Simons terms that are   required
for some groups $\cal G$. The relation between $D,q$ and $\cal G$ is  
summarised in Table I. The actions Eq.(\ref{oxid}) include the theories of pure  
gravity for
${\cal G}=A_{D-3}$, the bosonic sector of the M-theory effective action   for 
${\cal G}=E_8$ and the bosonic string effective action for  ${\cal G}=D_{24}$.

\begin{center}
\begin{tabular}{||c|c|c||}
\hline $q$&$D$&$ {\cal G}$\\
\hline\hline
$0$&$D$&$A_{D-3}$\\
\hline
$1$&$D$& $B_{D-2}~{\rm and}~ D_{D-2}$\\
\hline
$q$&$4$& $C_{q+1}$\\
\hline
$1$&$8$& $E_6$\\
\hline
$1$&$9$& $E_7$\\
\hline
$0$&$11$& $E_8$\\
\hline
$0$&$5$& $G_2$\\
\hline
$1$&$6$& $F_4$\\
\hline
\end{tabular}

{\small   Table I : $q$ and $D$ for each simple Lie group  
$G$ in the reduced theory.}
\end{center}  

\vskip .3cm
\hskip .4cm\epsfbox{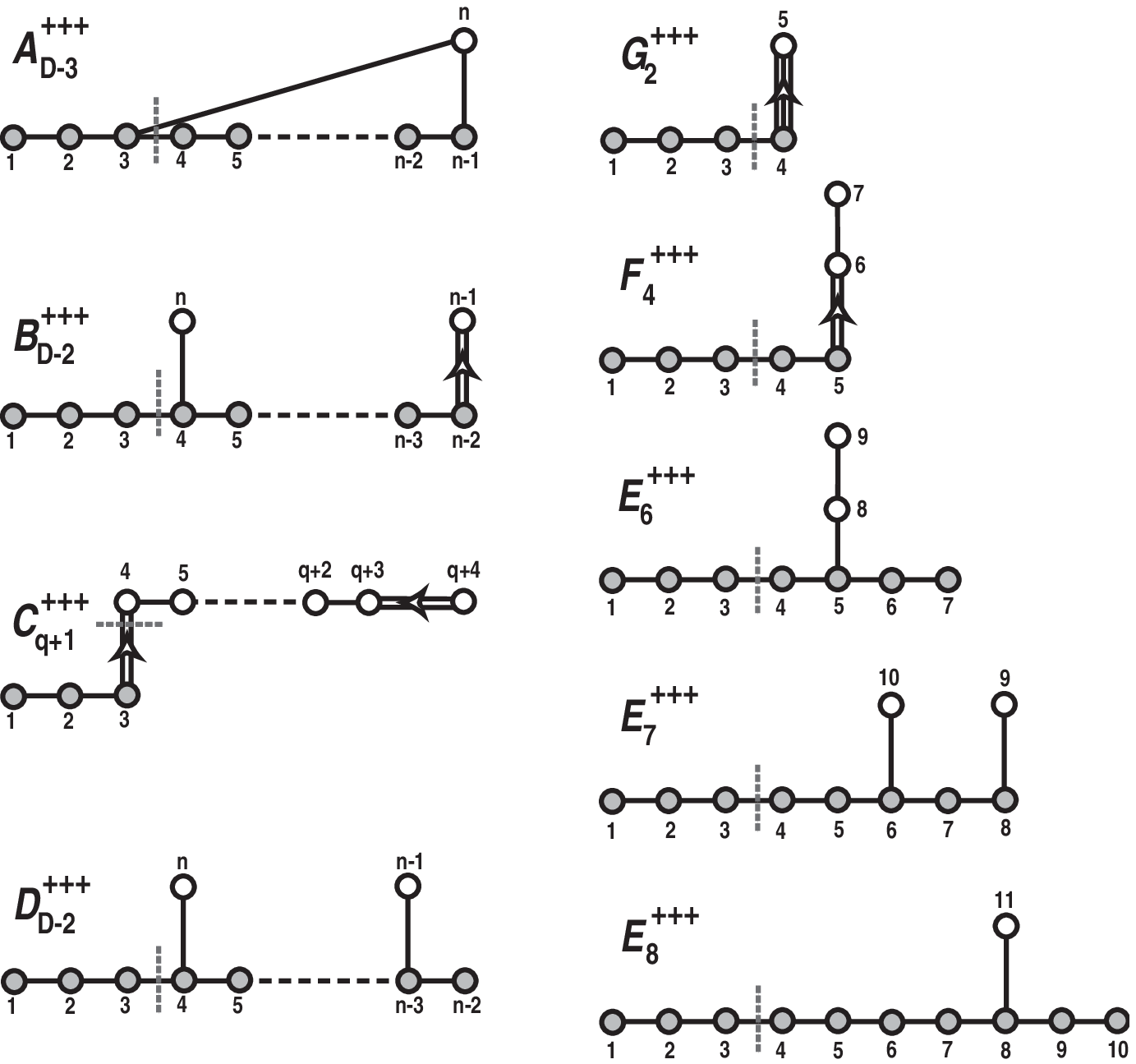}
\begin{quote}\begin{center}
\baselineskip 10pt {\small Fig.1.  Dynkin diagram of
$\G$.}
\end{center} {\small The nodes of the gravity line are
shaded. The   Dynkin diagram of $\cal G$ is that part of the diagram of $\G$ 
which sits on   the right of the dashed line. The first three nodes define the
Kac-Moody extensions.}
\end{quote} 

We consider the  real form of the symmetrisable Lorentzian Kac-Moody   algebra
$\G$ defined from the algebra of
$\cal G$ by extending its Dynkin diagram with three nodes in a line
\cite{olivegw02}, the first one being the affine node extending $\cal G$ to its
affine Kac-Moody extension\footnote{Throughout the paper, we use the same
notation for   groups and  Lie algebras.}.
The Dynkin diagram of $\G$ contains the diagram, hereafter  
referred to as  {\em  the gravity line}, of the subalgebra
$A_{D-1}=SL(D)$. The algebra of $GL(D)\times U(1)^q$, which has the   same rank
as
$\G$, is always a subalgebra of $\G$. The Dynkin diagrams of all simple  
$\G$ are depicted in Fig.1.

We want ultimately  $\G$ to fully characterise the symmetry of the    action $S$
defined by Eq.(\ref{oxid}), or of some more general theory. As an   enlargement
of the {\em global} symmetry group $\cal G$ arising in the dimensional  
reduction of
$S$,
$\G$ should also define  a {\em global} symmetry. This poses a dilemma.   The
metric and the form field strengths  in
$S$ are genuine space-time  fields and $S$ is invariant under the
{\em local} diffeomorphism and gauge groups.
How could a global symmetry encompass a local symmetry? The present   analysis
is an attempt to solve this dilemma by taking advantage of the infinite
dimensionality
  of the algebra
$\G$. More precisely we shall replace the action
$S$ by an action $\cal S$ explicitly invariant under the global $\G$   symmetry.
$\cal S$ will contain an infinite number of objects that are tensors   under
$SL(D)$. These will comprise  a symmetric tensor $g_{\mu\nu}$,  scalars
$\phi^u$ and  ($p_I -1$)-form potentials $A_{\mu_1\mu_2 \dots  
\mu_{p_I-1}} $ which shall be interpreted as the corresponding fields 
occurring in   Eq.(\ref {oxid})   taken at a {\em fixed} space time point.
Their motion in   space-time, as well as those of possible additional fields,
is expected to take place   through an infinite number of field derivatives at
this point,  encoded in other objects in
$\cal S $.

We recall that the algebra of ${\cal G}^{+++}$, whose rank $r=D+q$, is  
entirely defined by the commutation relations of its Chevalley generators and
by   the Serre relations. Let
$E_m,F_m$ and
$H_m,
\ m=1,2,\dots r$~, be the  generators satisfying
\begin{eqnarray}
   [H_m,H_n]=0~~, ~~[H_m,E_n]&=&A_{mn}E_n \, ,\nonumber \\
\label{chevalley}
   [H_m,F_n]&=&-A_{mn} F_n~~,~~[E_m,F_n]=\delta_{mn} H_m\, ,
\end{eqnarray}  where $A_{mn}$ is the Cartan matrix. The Cartan     subalgebra
is generated by $H_m$, while  the positive (negative) step operators  are   
  the  $E_m$ ($F_n$) and their multi-commutators subject to the Serre relations
\begin{equation}
\label{serre} [E_m,[E_m,\dots,[E_m,E_n]\dots]]=0,\quad
[F_m,[F_m,\dots,[F_m,F_n]\dots]]=0\, ,
\end{equation} where the number of $E_m$ ($F_m$) acting on $E_n$     ($F_n$) is
given by
$1-A_{mn}$.

The generators of the $GL(D)$ subalgebra are taken to be $K^a{}_b\  
(a,b=1,2,\ldots ,D)$   with commutation relations
\begin{equation}
\label{Kcom} [K^a_{~b},K^c_{~d}]   =\delta^c_b   K^a_{~d}-\delta^a_dK^c_{~b}\, 
.
\end{equation} The Cartan generators $H_m$ in the Chevalley basis are linear  
combinations of the
$K^a{}_a$ and of the
  abelian generators $R_u \ (u=1,2,\dots,q)$. The Cartan generators  
corresponding to the gravity line are given by $H_m=K^m_{~m}-K^{m+1}_{~m+1}$
with $m=1  
\dots D-1$. We shall  see how to express  the positive  and negative step
operators $E_\alpha$ and  $(   E^T_{-\alpha})$ as $SL(D)$ tensors. Here
$E_\alpha$ designate the $E_m$ and all the generators obtained from them by
multi-commutators, up to a multiplicative normalisation factor.
$E^T_{-\alpha}$ designate the generator obtained by substituting in  
$E_\alpha$ all
$E_m$ by $F_m$ and taking the multi-commutators in reverse order. Their
normalisation factor is chosen so that the  invariant bilinear form in  
$\G$ \cite{kac83} is normed to
\begin{equation}
\label {norm}
\langle E_{\alpha\, s},  E^T_{-\beta\, t}\rangle=  \delta_{st}\,
\delta_{\alpha\beta}\, ,
\end{equation} where $s$ and $t$ label possible degeneracies of the   root
$\alpha$.

  The positive (negative) step operators in the   
$A_{D-1}$ subalgebra are, from Eq.(\ref{Kcom}), the
  $K^a{}_b$ with $b>a$ ($b<a$).  They define the level zero step   operators of
the
${\cal G}^{+++}$ adjoint representation. The positive (negative) levels   of the
adjoint representation of ${\cal G}^{+++}$ are defined as follows. One   takes
a set of
$q$ non-negative (non-positive)   integers, excluding $q$ zeros, where  
$q$ is the number of  simple roots of ${\cal G}^{+++}$ not contained in the  
gravity line. The
$q$ integers count the number of times each such root appears in the
decomposition of the adjoint representation of ${\cal G}^{+++}$ into  
irreducible representations of  $A_{D-1}$. Positive (negative) levels contain 
only   positive (negative) roots and the number of irreducible representations
of
$A_{D-1}$ at each level is finite.  All step operators may be written as
irreducible tensors of the $A_{D-1}$ subalgebra of ${\cal G}^{+++}$.   Their
symmetry properties are fixed by the Young tableaux describing the   irreducible
representations appearing at a given level. Iterative procedures to   compute
the step operators at any  level can be devised. They build, together with  
the Cartan generators
$K^a{}_a\ (a=1,2,\ldots ,D)$  and  $R_u \ (u=1,2,\dots,q)$, the full   content
of the adjoint representation  of ${\cal G}^{+++}$.

The commutators of all step operators are generated by the commutators   of
  step operators  corresponding to simple roots. At level zero these   `simple
step operators' are, from Eq.(\ref{Kcom}), the  $K^a{}_{a+1}\  
(a=1,2,\dots,D-1)$. Those not contained in the gravity line are components of
tensors occurring   at low levels. Tensor transformation properties are given by
\begin{equation}
\label{tensor} [K^a_{~b}, R^{\quad c_1\dots c_r}_{ d_1\dots  
d_s}]=\delta^{c_1}_b R^{\quad a\dots c_r}_{ d_1\dots d_s}+\dots+\delta^{c_r}_b
R^{\quad   {c_1}\dots a}_{ d_1\dots d_s}-\delta_{d_1}^a R^{\quad {c_1}\dots
c_r}_{b\dots   d_s}-\dots -\delta_{d_s}^a R^{\quad {c_1}\dots c_r}_{ d_1\dots
b}\, .
\end{equation} For instance, we see from the $E_8^{+++}$ Dynkin diagram   in
Fig.1, which characterises M-theory, that the only non-gravitational simple  
step operator occurs at level one and is associated to the root   
$\alpha_{11}$. The only
$A_{10}$ representation at that level is  a third rank antisymmetric   tensor
$R^{abc}$. This is of course to be expected as the `electric' root  
$\alpha_{11}$ is generated in the dimensional reduction by a 3-form potential
in the   action Eq.(\ref{oxid}). One has thus
\begin{equation}
\label{root} [K^a_{~b},R^{efg}]   =\delta^e_b R^{afg}+\delta^f_b
R^{eag}+\delta^g_b R^{efa}\,  ,
\end{equation} and from the Chevalley relations Eq.(\ref{chevalley})   one
easily identifies the simple step operator at level one as
  $R^{9\,10\, 11}$.  In general, when $q$ dilatons are present in the   action
$S$, the rank of
$\G$ is $D+q$. The $q$  abelian generators
$R_u$ of its subgroup $GL(D)\times U(1)^q$  have non vanishing   commutators
with the tensor step operators
$R^{a_1a_2\dots a_r}$ associated to electric or magnetic  simple roots.   In
dimensional reduction, these arise from
  $r$-form potentials where $r=p_I-1$ for an electric root and  
$r=D-p_I-1$ for a magnetic one. We can read off their commutation relation, in
the   normalisation given for the dilaton in Eq.(\ref{oxid}), namely
\begin{equation}
\label{dilaton} [R_u,R^{a_1a_2\dots a_r}]=
-\varepsilon\frac{a^u_I}{2}R^{a_1a_2\dots a_r}\, ,
\end{equation} where $\varepsilon=+1$ for an electric root and $-1$ for   a
magnetic one \cite{ehtw}. Eqs.(\ref{Kcom}), (\ref{tensor}) and (\ref{dilaton})
allow to express all step operators associated to the simple roots of $\G$ and
their   commutations relations with the generators of the Cartan subalgebra as
$A_{D-1}$   tensors. These are listed in Appendix A for all simple  algebra
$\G$.  By multiple   commutators and reduction at any level into suitably
normed irreducible $A_{D-1}$   tensors, one may in principle, recursively,
list the positive step operators at any   level.

To switch from positive $K^a{}_b\, (b>a)$ step operators to negative   ones it
suffices to interchange  upper and lower indices.  Writing the negative   of
$R^{\quad c_1\dots c_r}_{ d_1\dots d_s}$  as $R_{\quad c_1\dots c_r}^{  
d_1\dots d_s}$, one  verifies that the correct tensor transformations of the  
negative step operators follows from those of the positive ones by writing 
the commutator Eq.(\ref{tensor}) 
in   reverse order, namely
$[R_{\quad c_1\dots c_r}^{ d_1\dots d_s}, K^b{}_a]$.  One may then    form  all
negative step operators by writing  the multi-commutators of the simple   roots
defining the positive step operators, interchanging in each simple step  
operator upper and lower indices, and taking the multi-commutators in reverse  
order. In this way the  bilinear form Eq.(\ref{norm}) becomes
\begin{equation}
\label{normalisation}
\langle R^{\quad a_1\dots a_r}_{ b_1\dots b_s}\, R_{\quad d_1\dots   d_r}^{
c_1\dots c_s}\rangle
=\delta^{c_1}_{b_1}\dots\delta^{c_s}_{b_s}\delta^{a_1}_{d_1}\dots\delta^ 
{a_r}_{d_r}\, . 
\end{equation}

\subsection{The temporal involution and the coset space ${\G}/K^{+++} $}

  The metric $g_{\mu\nu}$ at a fixed space time point  parametrises the   coset
$GL(D)/SO(D-1,1)$. To construct a $\G$ invariant action $\cal S$   containing
such a tensor, we shall build   a non linear realisation of $\G$ in a coset  
space
${\G}/K^{+++}$ where the subgroup $K^{+++}$ contains the Lorentz group
$SO(D-1,1)$.  We use a recursive construction based on the level  
decomposition of
$\G$. As at each level the $SO(D-1,1)$ invariance must be realised for   a
finite number of generators,  we cannot use the Chevalley involution to build  
the coset
${\G}/K^{+++}$ . Rather we shall use a `temporal'  involution  from   which  
the required non-compact generators of $K^{+++}$ can be selected.

The  `temporal'  involution is defined in the following way. For the  
generators of the Cartan subalgebra we take, as in the Cartan involution,
\begin{equation}
\label{cartanmap} K^a{}_a \mapsto -K^a{}_a\qquad R_u\mapsto -R_u\, ,
\end{equation} or equivalently in the Chevalley basis
\begin{equation}
\label{map} H_m \mapsto -H_m\,  ,
\end{equation} The simple step operators are mapped according to
\begin{equation}
\label{stepmap} E_m \mapsto -\epsilon_m F_m\qquad F_m \mapsto   -\epsilon_m
F_m\, .
\end{equation} Here   $E_m$  is expressed as a $A_{D-1}$ tensor  and  
$F_m$ as the tensor with upper and lower indices interchanged. $\epsilon_m$  
is defined as
$+1$ if the number of `1' indices (that is the number of time indices)   is
even and (-1) otherwise. It is straightforward to verify that the Chevalley  
presentation Eq.(\ref{chevalley}) is preserved under the map given in  
Eqs.(\ref{map}) and (\ref{stepmap}).

Any positive step operator $E_{\alpha\, s}$ is expressed (up to a  
normalisation constant) as
\begin{equation} E_{\alpha\, s}= [E_q,[E_{q-1},\dots,[E_2,E_1]\dots]]\,   .
\end{equation} Under the map Eq.(\ref{stepmap}) one gets
\begin{eqnarray}
\label{transpose} E_{\alpha\, s}&\mapsto &(-1)^q\,  
\epsilon_1\epsilon_2\dots
\epsilon_{q-1}\epsilon_q\,    [F_q,[F_{q-1},\dots,[F_2,F_1]\dots]]\nonumber\\
&&=(-1)\, \epsilon_1\epsilon_2\dots \epsilon_{q-1}\epsilon_q\,  [[\dots
[F_1,F_2],\dots,F_{q-1}],F_q]\nonumber\\ &&=-\epsilon_\alpha \,   E^T_{-\alpha\,
s}\, .
\end{eqnarray} As seen from the previous discussion,  the negative    step
operator
$ E^T_{-\alpha\, s}$ is obtained by interchanging upper and lower   indices in
$E_{\alpha\, s}$. The factor
$\epsilon_\alpha$ is $\pm1$, according to the parity of the number of   time
indices occurring in
$E_{\alpha\, s}$ (or in $ E^T_{-\alpha}$). Clearly all commutation   relations
in $\G$ are preserved under the mapping Eqs.(\ref{map}) and (\ref{stepmap}),  
and so is its bilinear form, as seen from Eq.(\ref{norm}) and
(\ref{normalisation}).   This mapping constitutes an involution that we label
the temporal involution. We   define the subgroup $ K^{+++}$ of $\G$ as the
subgroup invariant under this involution. Its generators are
\begin{equation}
\label {subgroup} E_{\alpha\, s}  -\epsilon_\alpha \, E^T_{-\alpha\,   s}\, .
\end{equation}
  $ K^{+++}$ contains the Lorentz group $SO(D-1,1)$ and all generators   with
$\epsilon_\alpha =-1$ are non-compact \footnote{The occurrence of a  
non-compact $K^{+++}$ invariant under the temporal involution is the infinite
dimensional   counterpart of what is happening in dimensional reduction when
one compactifies the   time. Indeed, for instance, 11 dimensional supergravity
compactified down to three    spatial dimensions gives a coset $G / H$ with
$G=E_8$ and $H=SO^*(16)$ with $SO^*(16)$   being a non-compact form of $SO(16)$
with maximal compact subgroup $U(8)$ \cite{hullj98}.}.

\subsection{Non-linear realisation of $\G$ in ${\G}/K^{+++} $ }

We will follow a similar line of thought as the one  developed in reference
\cite{damourhn02} in the context of $E_8^{++}$. Consider a group element $\cal
V$  built out of Cartan and positive step operators in $\G$. It takes the form
\begin{equation}
\label{positive} {\cal V}= \exp (\sum_{a\ge b} h_b^{~a}K^b_{~a} -  
\sum_{u=1}^q
\phi^u R_u) \exp (\sum
\frac{1}{r!s!} A^{\quad a_1\dots a_r}_{ b_1\dots b_s} R_{\quad a_1\dots   a_r}^{
b_1\dots b_s} +\cdots)\, .
\end{equation} We have written it so that the first exponential contains only 
level zero   operators (i.e the Cartan and the level zero positive step
operators) and the second one contains the positive step operators of level
strictly greater than zero. The tensors $ h_b^{~a}, \phi^u, A^{\quad a_1\dots
a_r}_{ b_1\dots   b_s}$, bear a priori no relation with the metric, the dilaton
and the potentials  of   the $p_I$ form field strengths 
$F_{p_I}$ entering the action Eq.(\ref{oxid}). However we shall see   that a
dictionary can be established relating the  tensors which appear at low   levels
with the fields occurring in  Eq.(\ref{oxid}) at a {\em fixed}   space-time
point. For higher levels the dictionary between group parameters and
space-time   fields should arise, as discussed in Section 3 , from the analysis
of the   dynamics encoded in the
$\G$ invariant action
$\cal S$ below.

A differential motion in the coset ${\G}/K^{+++}$ can be constructed   from
Eq.(\ref{positive}). Define
\begin{equation}
\label {sym} dv= d{\cal V} {\cal V}^{-1}\qquad d\tilde v=  
\widetilde{\cal V}^{-1}d\widetilde {\cal V}
\qquad;\qquad dv_{sym}=\frac{1}{2} (dv+d\tilde v)\, .
\end{equation} Here $\widetilde{\cal V}$ is obtained from ${\cal V}$ by   the 
map obtained by flipping the sign in the RHS of Eqs.(\ref {map}) and
(\ref{transpose}), namely $H_m \mapsto + H_m$ and
$E_{\alpha\, s}
\mapsto +\epsilon_\alpha \, E^T_{-\alpha\, s}$. As $dv$ and $d\tilde v$   are
differentials in the Lie algebra,  $dv_{sym}$ contains only the Cartan  
generators and the combinations of step operators $E_{\alpha\, s}   
+\epsilon_\alpha \, E^T_{-\alpha\, s}$. Hence  it defines a differential motion
in the   coset ${\G}/K^{+++}
$.  This linear combination of  step operators  takes the form
$ R^{\quad c_1\dots c_r}_{ d_1\dots d_s} +
\epsilon_{c_1}\dots\epsilon_{c_r}\epsilon_{d_1}\dots\epsilon_{d_s}
  R_{\quad c_1\dots c_r}^{ d_1\dots d_s}$ where $\epsilon_a =-1$ if $a=1$ and
$\epsilon_a=+1$ otherwise.

To construct the action $\cal S$ we wish  to map  a manifold
$\cal M$ into
$\G$. We do not want to take for
$\cal M$ a space-time manifold, as this might require the explicit  
introduction of local symmetries which we hope to be hidden in the infinite
algebra   of
$\G$.  We shall take for $\cal M$  a one-dimensional world-line in $\xi$-space,
i.e.
$dv_{sym}= v_{sym}(\xi)d\xi$. Here no connection is imposed a priori between  
$\xi$-space and space-time.

A reparametrisation invariant action is then
\begin{equation}
\label{actionG} {\cal S}=\int d\xi  \frac{1}{n(\xi)}\langle  
v_{sym}^2(\xi)\rangle=
\frac {1}{4}\int d\xi
\frac{1}{n(\xi)}\left\langle
\,
\left(\frac{d{\cal V(\xi)}}{d\xi} {\cal V(\xi)}^{-1}+\widetilde{\cal  
V}^{-1}\frac {d\widetilde {\cal V}}{d\xi}\right)^2\,\right
\rangle ,
\end{equation} where ${\cal V(\xi)}$ are the group parameters appearing   in
Eq.(\ref{positive}) that are now fields dependent on the  variable  
$\xi$, and
$n(\xi)$ is an arbitrary lapse function ensuring reparametrisation   invariance
on the world-line. The `trace' $\langle~\rangle$ means the invariant   bilinear
form on
$\G$ \cite{kac83}  which can in principle be computed in the recursive  
approach. It ensures the invariance of the non-linear action $\cal S$ defined
on the coset space
${\G}/K^{+++} $ under global $\G$ transformations.

We now compute the level zero of the action Eq.(\ref{actionG}), that is   the
terms generated by  $K^b_{~a}~(a\ge b)$ and $R_u$ in $\cal V$. Writing $dv  
=v(\xi)d\xi$, one obtains from Eqs.(\ref{positive}) and (\ref{sym}) the
contribution   of the level zero to $v_{sym}(\xi)$,
\begin{equation}
\label{v0} v_{sym}^0(\xi)=-\frac{1}{2}\sum_{a\ge b} [e^  h (\frac{de^{-
h}}{d\xi})]_b^{~a} (K^b_{~a}+
\epsilon_a
\epsilon_b K^{a}_{~b}) -\sum_{u=1}^q
\frac{d\phi^u}{d\xi}  R_u\, ,
\end{equation} where $ h$ is triangular matrix with elements   
$h_b^{~a}$. We now evaluate $\langle (v_{sym}^0)^2\rangle$. The invariant form
in $\G$ for   the Cartan generators  is given by \cite{ehtw}
\begin{equation}
\label{killing}
\langle K_a^a K_b^b\rangle =G_{ab}\quad{\rm with} \quad  G= I_D -
\frac{1}{2}\Xi_D\quad,\quad \langle R_u R_v\rangle   =\frac{1}{2}\delta_{uv}\, ,
\end{equation} where $\Xi_D$ is a D-dimensional matrix with all entries   equal
to unity. For the step operators we have
\begin{equation}
\label{killingstep}
\langle K^b_{~a}K_{~c}^d\rangle= \delta_c^b\delta_a^d \qquad a>b\,.
\end{equation} The $\epsilon$-symbols defining the temporal involution   allow
the raising or lowering of the
$a,b$ indices of the  
$\xi$-fields multiplying the negative step operator in
$\langle v_{sym}^0(\xi)v_{sym}^0(\xi)\rangle$ with the Minkowskian metric
$\eta_{ab}$ . This ensures that this   expression is a Lorentz scalar. We get
thus the Lorentz invariant action at level   zero, ${\cal S}^0$, using
Eqs.(\ref{killing}), (\ref{killingstep}) and the   triangular structure of
$h$,
\begin{eqnarray}
\label{sym2} {\cal S}^{(0)}&=&\int d\xi
\frac{1}{n(\xi)}\langle(v_{sym}^0(\xi))^2\rangle\, ,
\nonumber\\ &=&\frac{1}{2}\int d\xi
\frac{1}{n(\xi)}\left\{  [e^  h (\frac{de^{-  h}}{d\xi})]_b^{~a} [e^  h  
(\frac{de^{- h}}{d\xi})]^{T~b}_a +[e^  h (\frac{de^{-  h}}{d\xi})]_b^{~a} [e^ 
h   (\frac{de^{-
h}}{d\xi})]_a^{~b}\right.\nonumber\\&&\qquad\qquad\qquad\left.- ([e^  h
(\frac{de^{-  h}}{d\xi})]_a^{~a})^2 +
\sum_{u=1}^q (\frac{d\phi^u}{d\xi})^2\right\}  ,
\end{eqnarray} where the summation is performed over Lorentz indices.   Note
that the lower indices of
$e^{-h}$ and the upper indices of $e^h$ cannot be lowered or raised by   the
Lorentz metric. To avoid confusion we label these indices with greek letters,  
namely we define `vielbein'
\begin{equation}
\label{vielbein}
  e_\mu^{~a}=(e^{-h})_\mu^{~a}\quad e^{~\mu}_b=(e^h)^{~\mu}_b\quad;\quad
g_{\mu\nu} =e_\mu^{~a}e_\nu^{~b}\eta_{ab}\, .
\end{equation} Although we have not  yet introduced a space-time, we   shall
name the $a$ indices  flat  and the $\mu$ indices  curved. As a result   of the
temporal involution and of the scalar product $\langle~\rangle$\ in  
$\G$, the flat-index tensors have been endowed with a Lorentz metric while  
curved-index tensors define a metric in $GL(D)/SO(D-1,1)$. Hence, for any
$\xi$, we   are allowed to identify
$g_{\mu\nu}(\xi)$ in Eq.(\ref{vielbein}) as the metric tensor in $S$,  
Eq.(\ref{oxid}), at a fixed space-time point.

Using Eq.(\ref{vielbein}), one can  rewrite the action Eq.(\ref{sym2})   with
flat or curved indices as
\begin{eqnarray}
\label{faction0} {\cal S}^{(0)}&=&\frac{1}{2}\int d\xi
\frac{1}{n(\xi)}\left[  
e^{~\mu}_b\frac{de_\mu^{~a}}{d\xi}(\frac{de_{a\nu}}{d\xi} e^{\nu
b}+e^{~\nu}_a\frac{de_\nu^{~b}}{d\xi})- (e^{~\mu}_b\frac{de_\mu^{~b}}{d\xi})^2 +
\sum_{u=1}^q (\frac{d\phi^u}{d\xi})^2 \right]\\ &\rm or&\nonumber\\
\label{action0} {\cal S}^{(0)}&=&\frac{1}{2}\int d\xi
\frac{1}{n(\xi)}\left[\frac{1}{2}(g^{\mu\nu}g^{\sigma\tau}- 
\frac{1}{2}g^{\mu\sigma}g^{\nu\tau})\frac{dg_{\mu\sigma}}{d\xi}
\frac{dg_{\nu\tau}}{d\xi}+\sum_{u=1}^q
\frac{d\phi^u}{d\xi}\frac{d\phi^u}{d\xi}\right].
\end{eqnarray} At higher levels, the tensors multiplying the step   operators
couple nonlinearly to the level zero objects and  between themselves.   The
coupling to the metric and to $\phi^u$ can be  formally written down   for all
levels, but the self-coupling of the $A^{\quad a_1\dots a_r}_{ b_1\dots b_s}  
$depend specifically on the group
$\cal G$.

  Consider a general
$A_{D-1}$ tensor $A_{a_1\dots a_r}^{\quad b_1\dots b_s}$ parametrising a
normalised step operator
$R^{\quad a_1\dots a_r}_{ b_1\dots b_s}$. The commutation relations of  
$R^{\quad a_1\dots a_r}_{ b_1\dots b_s}$ with the $K^b_{~a}$ are given by the  
tensor transformations as in Eq.(\ref{Kcom}) and (\ref{tensor}) and those with  
$R_u$ have the form
\begin{equation}
\label{gendilaton} [R_u,R^{\quad a_1\dots a_r}_{ b_1\dots b_s}]=  
\lambda_u\, R^{\quad a_1\dots a_r}_{ b_1\dots b_s}\, .
\end{equation} Here
$\lambda_u=\sum\lambda_{u,i}$ where the $\lambda_{u,i}$ are the scale
parameters of the simple step operators entering the multiple   commutators
defining $R^{\quad a_1\dots a_r}_{ b_1\dots b_s}$. This property   follows 
from the Jacobi identity. Identifying for simple step operators
$\lambda_{u,i}$   with
$-\varepsilon a^u_I/2$ in Eq.(\ref{dilaton}) we may  identify for any  
$\xi$ the
$\phi^u(\xi)$ in $\cal S$, Eq.(\ref{actionG}), with the dilatons fields   in
$S$, Eq.(\ref{oxid}),  at a fixed space time point. The particular
$A_{a_1\dots a_r}^{\quad b_1\dots b_s}(\xi)$ multiplying the step   operators
belonging to the subgroup $\cal G$ can be similarly identified to the
corresponding potential forms in $S$ along with their duals.

It is  straightforward to compute the contribution
$v^{(A)}$ to
$v$ of a given tensor when commutators of the $R^{\quad a_1\dots a_r}_{  
b_1\dots b_s}$ between themselves are disregarded. On gets
\begin{equation}
\label{vform} v^{(A)}=\frac{1}{r!s!}dA_{\mu_1\dots \mu_r}^{\quad  
\nu_1\dots
\nu_s} \, \exp (-\sum_{u=1}^q
\lambda^u \phi^u)\, e^{~\mu_1}_{{a}_1}\dots  
e^{~\mu_r}_{{a}_r}e_{\nu_1}^{~{b}_1}
\dots e_{\nu_s}^{~{b}_s}\, R^{\quad a_1\dots a_r}_{ b_1\dots b_s}\,  .
\end{equation} The contribution ${\cal S}^{(A)}_0$ of $v^{(A)}$ to the   action
${\cal S}$ is computed as previously and one gets
\begin{equation}
\label{actionA} {\cal S}^{(A)}_0=\frac{1}{2}\int d\xi
\frac{1}{n(\xi)}\left[\frac{1}{r!s!}\exp (-\sum_{u=1}^q 2\lambda^u  
\phi^u)
\frac{dA_{\mu_1\dots \mu_r}^{\quad \nu_1\dots
\nu_s}}{d\xi} g^{\mu_1{\mu}^\prime_1}...\,
g^{\mu_r{\mu}^\prime_r}g_{\nu_1{\nu}^\prime_1}...\,   g_{\nu_s{\nu}^\prime_s}
\frac{dA_{{\mu}^\prime_1\dots {\mu}^\prime_r}^{\quad {\nu}^\prime_1\dots
{\nu}^\prime_s}}{d\xi}\right].
\end{equation} The full action can only be approached in a recursive   way. In
${\cal S}^{(A)}_0$, one must replace derivatives by non linear generalisations  
to take into account the non vanishing commutators between tensor step  
operators. We represent such terms by `covariant' derivatives symbol
$D/D\xi$. There evaluation is group dependent. As an example we give the
covariant derivative of the $6$-form appearing at level 2 in
$E_8^{+++}$. One has, as in $E_8^{++}$ \cite{damourhn02},
\begin{equation}
\label{e8}
\frac{D}{D\xi}A_{a_1a_2a_3a_4a_5a_6}=\frac{d}{d\xi}A_{a_1a_2a_3a_4a_5a_6  }+ 10
A_{[a_1a_2a_3}\frac{d}{d\xi} A_{a_4a_5a_6 ]}\, .
\end{equation} Formally the full action $\cal S$ is
\begin{equation}
\label{full} {\cal S}={\cal S}^{(0)}+\sum_A{\cal S}^{(A)}\, ,
\end{equation}
$${\cal S}^{(A)}=\frac{1}{2}\int d\xi
\frac{1}{n(\xi)}\left[\frac{1}{r!s!}\exp (-\sum_{u=1}^q 2\lambda^u  
\phi^u)
\frac{DA_{\mu_1\dots \mu_r}^{\quad \nu_1\dots
\nu_s}}{d\xi} g^{\mu_1{\mu}^\prime_1}...\,
g^{\mu_r{\mu}^\prime_r}g_{\nu_1{\nu}^\prime_1}...\,    g_{\nu_s{\nu}^\prime_s}
\frac{DA_{{\mu}^\prime_1\dots {\mu}^\prime_r}^{\quad {\nu}^\prime_1\dots
{\nu}^\prime_s}}{d\xi}\right] $$ where the sum on $A$ is a summation   over all
tensors appearing at all positive levels in the decomposition of $\G$   into
irreducible representations of $A_{D-1}$.

One may expand $\cal S$ given in Eq.(\ref {full}) in power of fields  
parametrising the positive step operators. Up to quadratic terms, the result
${\cal   S}^{(Q)}$ is obtained by retaining in
$v(\xi)$ terms independent or linear in these fields. Define the   one-forms
$[dA]$ and the moduli $p^{(a)}$ (or $p^{(\mu)}$), as in \cite{ehtw,ehw},
by\footnote{We can indifferently label the moduli by a curved or a flat index,
as it is uniquely defined by the   diagonal vielbein in the triangular gauge
once the $g_{\mu\nu}$ have been   chosen. The
  position of this index as a subscript or superscript is then a matter   of
convention and has no tensor significance.}
\begin{eqnarray}
\label{flatform}
\exp (-\sum_{u=1}^q
\lambda^u \phi^u)\,
\hat e^{~\mu_1}_{{a}_1}\dots\hat e^{~\mu_r}_{{a}_r}\hat   e_{\nu_1}^{~{b}_1}
\dots\hat e_{\nu_s}^{~{b}_s}\,\frac{dA_{\mu_1\dots \mu_r}^{\quad  
\nu_1\dots
\nu_s}}{d\xi}&\stackrel{def}{=}&\frac{ [dA]_{a_1\dots a_r}^{\quad   b_1\dots
b_s}}{ d\xi}\, ,\\
\label{flatmetric}
\hat e^{~\mu}_b\frac{d\hat e_{\mu}^{~a}}{d\xi}&\stackrel{def}{=}&\frac   {d
p^{(a)}}{d\xi}\, ,
\end{eqnarray} where   $\hat e$ means that only the diagonal vielbein   are
kept. Taking into account that in Eq.(\ref{faction0}), only the first term in  
the right hand side contains non diagonal elements of the vielbein, we get
\begin{eqnarray}
\label{linear} {\cal S}^{(Q)}&=&\int d\xi\frac{1}{n(\xi)}  
\left[\sum_{a=1}^D (\frac{dp^{(a)}}{d\xi})^2-\frac{1}{2}(\sum_{a=1}^D  
\frac{dp^{(a)}}{d\xi})^2 +\frac{1}{2}\sum_{u=1}^q
(\frac{d\phi^u}{d\xi})^2\right.\nonumber \\
&&\left.+\frac{1}{2}(e^{~\mu}_b\frac{d e_\mu^{~a}}{d\xi}\frac{d  
e_{a\nu}}{d\xi}
  e^{\nu b})^{(1)}+\frac{1}{2}\frac{1}{r!s!}\sum_A\frac{[dA]_{a_1\dots  
a_r}^{\quad b_1\dots b_s}}{ d\xi}\, \frac{[dA]^{\quad a_1\dots a_r}_{ b_1\dots  
b_s}}{ d\xi}\right]  ,
\end{eqnarray} where the superscript (1) in the vielbein term indicates that  
only terms quadratic in $\, h_b{}^a\, (a>b)$ are kept.  In the next section we  
shall produce solutions of the  action Eq.(\ref{linear})  which are {\em
exact}   solutions of the full action Eq.(\ref {full}).

\setcounter{equation}{0}
\section{BPS states in  $\G$ }
\subsection{Extremal brane       solutions of the oxidised actions $S$}   We
first recall how   closed extremal branes   are obtained  as solutions of  
general relativity coupled to forms in the oxidised theories described by the  
actions Eq.(\ref {oxid}).  For simplicity we restrict ourselves to one
dilaton.   One considers the diagonal metrics
\begin{equation} ds^2=-e^{2p^{(1)}} d\tau^2+\sum_{\mu=2}^{D-p}   e^{2p^{(\mu)}}
(dx^\mu)^2+\sum_{\lambda=1}^pe^{2p^{(D-p+\lambda)}}(dy^\lambda)^2\, ,
  \label{metric}
\end{equation} where $y^\lambda$  label the $p$  compact coordinates.   The
functions $p^{(\alpha)}~ (\alpha=1,2,\dots D)$ depend only on the    transverse
coordinates $x^\mu$ in the non-compact dimensions and  allow for   multi-centre
solutions. We choose $p\ge q_A$ where $q_A$ is the dimension of the   brane. If
$q_A<p$, we take  a lattice of  $q_A$-branes in the  compact directions  
transverse to the brane and average over them. Here $q_A$ designates either an  
electrically charged $q_E$-brane  with respect to the form
$p_I$-form field strength $F_{p_I}$, or its magnetic dual  $q_M$-brane.

One  choose the  ans\"atze for electric and magnetic branes
\begin{eqnarray}
\mbox{Electric}& :& A_{\tau
\lambda_1\dots\lambda_{q_E}}=\epsilon_{\tau
\lambda_1\dots\lambda_{q_E}}E_E(\{x^\nu\})\, ,
\label{electric}\\
\mbox{Magnetic}& :& \widetilde{A}_{\tau
\lambda_1\dots\lambda_{q_M}}=\epsilon_{\tau
\lambda_1\dots\lambda_{q_M}}
  E_M(\{x^\nu\})\, ,
\label{magnetic}
\end{eqnarray} where $ \widetilde{A}$ is the (magnetic) potential of   the dual
field strength
$\widetilde{F}$  defined by
\begin{equation}
\label{hodge}
\sqrt{-g} e^{a_I\phi}F^{\mu_1 \dots \mu_{p_I}}={1\over (D-p_I)!}
\epsilon^{\mu_1 \dots \mu_{p_I}\,  \nu_1 \dots \nu_{D-{p_I}}}
\widetilde{F}_{\nu_1 \dots \nu_{D-{p_I}}}\, .
\end{equation} Extremal  brane solutions of the coupled Einstein's   equations,
dilaton and forms are given by \cite{aeh}
\begin{eqnarray}
  p^{(1)}_A &=& - \frac{D-q_A-3}{\Delta}  \ln H_A\, ,
\nonumber\\
  p^{(\mu)}_A &= &  \frac{q_A+1}{\Delta} \ln H_A\, ,
\nonumber\\ p^{(D-p+\lambda)}_A&=& 
\frac{\delta^{(\lambda)}_A}{\Delta}\ln H_A\, , \nonumber\\
\phi_A &=& \frac{D-2}{\Delta}\varepsilon_A \, a \ln H_A\, ,
\label{phicond}
\end{eqnarray} where $H_A(\{x^\nu\})$ is a harmonic function related to   the
$E_A$ by $H_A=
\sqrt{2(D-2)/\Delta}\, E_A^{-1}$ and
\begin{equation}
\label{delta}
\Delta= (q_A+1)(D-q_A-3)+\frac{1}{2}a_A^2(D-2)
\end{equation}
  is invariant under electric-magnetic duality because $q_M=D-q_E-4$. In
Eq.(\ref{phicond})
$\delta^{(\lambda)}_A=-(D-q_A-3)$ or $(q_A+1)$ depending on wether
$y^\lambda$ is parallel or perpendicular to the
$q_A$-brane. The factor $\varepsilon_A$ is
$+1$  for  an electric brane and $-1$ for a magnetic one.  The harmonic  
functions
$H_A(\{x^\nu\})$ allow for parallel branes and are given  (for   
$D-p>3$) by
\begin{equation}
\label{branes} H_A=1+\sum_k \frac{Q_k}{|x^a-x^a_k|^{D-p-3}},  
\label{multicenter}
\end{equation} where the $x^a_k$ label the positions in non-compact space-time
of the branes with charge
$Q_k$.

The solutions Eq.(\ref{phicond}) satisfy \cite{ehw}
\begin{equation}
\label{embedding} (p+3-D)\,   p^{(\mu)}_A= p^{(1)}_A+ \sum_{\lambda=1}^p  
p^{(D-p+
\lambda)}_A\qquad
\mu=2,\dots, D-p\, ,
\end{equation} and, using Eq.(\ref{embedding}), one  verifies the   differential
relation
\begin{equation}
\label{extremal}
\sum_{\alpha=1}^D ( d p^{(\alpha)}_A)^2 -\frac{1}{2}(\sum_{\alpha=1}^D
dp^{(\alpha)}_A)^2 +
\frac{1}{2} (d\phi_A)^2= \frac{D-2}{\Delta}(d\ln H_A)^2\, .
\end{equation}

We will now show how the result Eqs.(\ref{extremal})  characterising   extremal
branes, that is BPS configurations, are related for any $\cal G$ to   exact
solutions of the $\G$ nonlinear action of Section 2. The group properties of  
these solutions, as well as their interpretation as dualities in a very general
context   encompassing the familiar properties of string and M-theory, will
emerge from  these exact solutions.

\subsection{BPS states as exact solutions of the $\G$ actions $\cal S$}

We shall look for solutions of the equations of motion derived from   
$\cal S$ and containing only one
$A(\xi)$-field,  or one non-diagonal $h(\xi)$-field,  with {\em given
indices}.   For such solutions, we may disregard all non-linearity in the
step operators.  We shall prove this statement for the $A$-fields and hence
also for the   non-diagonal
$h(\xi)$-field. The latter solutions will indeed follow  from the   former by
Weyl transformations. First consider the non-linear terms arising from arising
 the `covariant'
$D$-derivatives in   Eq.(\ref{full}). We label
$X$ the particular $A$-field component considered.  We  note that all terms in
$DA_{\mu_1\dots
\mu_r}^{\quad \nu_1\dots
\nu_s}$ are products of fields such that the sum of the levels of the   factors
in each term is equal to the level of $A_{\mu_1\dots \mu_r}^{\quad \nu_1\dots
\nu_s}$. From the ordering of factors in the  group element 
Eq.(\ref{positive}),   we see that all the fields contained in the `covariant'
derivatives are characterised by a level greater than zero. Consequently
potentially dangerous terms containing  more than one  $X$-factor  can not be
present in the covariant derivative of X. For such term to  appear in
any covariant derivative it must contain  a factor not equal to X.  We
can  thus satisfy  safely all the equations of motion of the fields   of level
greater than zero  by putting to zero all $A$-fields different from $X$. The
only term contributing to the equation   of motion for
$X$ comes then from $dX$. We still have to consider the equations of   motion
of the level zero fields, namely we have to check that the $h_b^{~a}$   with $
b<a$ can be put consistently to zero when only one $X$-field is non-zero. This
is the case as  in Eq.(\ref{actionA}) the term quadratic in $dX$ couples
    to  metrics, which are either expressible in terms of diagonal
vielbeins only or are at least quadratic in the
non-diagonal $h_b^{~a}$. Hence it is consistent to look for  solutions of the
equations of motion of   
$A(\xi)$ by replacing the action  $\cal S$ by its     simplified
version Eq.(\ref{linear}).

We shall as in Eqs.(\ref{electric}) and (\ref{magnetic}), consider $A$   to be
an antisymmetric tensor with a time index  $\tau$ and  
$r$ space indices coupled to a step operator of the  $\cal G$ subalgebra. The  
equation of motions are

\begin{itemize}

\item[]{\it a) The lapse constraint}.

Eq.(\ref{linear}), taking Eqs.(\ref{flatform}) and (\ref{flatmetric}) into
account, reads
\begin{equation}
\label{lapse}
\sum_{\alpha=1}^D   (\frac{dp^{(\alpha)}}{d\xi})^2-\frac{1}{2}(\sum_{\alpha=1}^D
\frac{dp^{(\alpha)}}{d\xi})^2 +\frac{1}{2} (\frac{d\phi}{d\xi})^2   
-\frac{1}{2} \exp [
\varepsilon a \phi -2p^{(\tau)}-2\sum_{\lambda=\lambda_1}^{\lambda_r}
p^{(\lambda)})] (\frac{dA_{\tau\lambda_1\dots \lambda_r}}{ d\xi})^2=0
\end{equation} Here we have taken one dilaton with scaling $\lambda
=-\varepsilon a/2$ in accordance with Eq.(\ref{dilaton}). Note that   this
relation is valid wether or not the  magnetic root is simple, as seen
in dimensional reduction. A crucial feature   of this equation is the minus sign
in front of the   exponential. Its origin can be traced back to the temporal 
involution defining our coset space, hence to Lorentz invariance, because
{\em both} magnetic  and electric potentials have a   time index.

\item[]{\it b)  The equation of motion for $A$}.

We take the lapse $n(\xi)=1$. One gets
\begin{equation}
\label {potential}
\frac{d}{ d\xi} \left( \, \exp [
\varepsilon a \phi -2p^{(\tau)}-2\sum_{\lambda=\lambda_1}^{\lambda_r}
p^{(\lambda)}]\frac{dA_{\tau\lambda_1\dots
\lambda_r}}{ d\xi} \, \right) =0\, .
\end{equation}

\item[]{\it c) The dilaton equation of motion}.
\begin{equation}
\label{dilat} -\frac{d^2\phi}{ d\xi^2}-\frac{1}{2}\varepsilon a\exp [
\varepsilon a \phi -2p^{(\tau)}-2\sum_{\lambda=\lambda_1}^{\lambda_r}
p^{(\lambda)}](\frac{dA_{\tau\lambda_1\dots
\lambda_r}}{ d\xi})^2=0\, .
\end{equation}

\item[]{\it d) The vielbein equations of motion}.
\hskip -1cm\begin{equation}
\label{vielbein1} -2\frac{d^2p^{(\alpha)}}{ d\xi^2} +\sum_{\beta=1}^D
\frac{d^2p^{(\beta)}}{ d\xi^2}=0\qquad
\alpha\neq \tau ,\lambda_i~~ (i=1,2\dots r)\, ,
\end{equation}
\begin{equation}
\label{vielbein2} -2\frac{d^2p^{(\alpha)}}{ d\xi^2} +\sum_{\beta=1}^D
\frac{d^2p^{(\beta)}}{ d\xi^2}+\exp [
\varepsilon a \phi -2p^{(\tau)}-2\sum_{\lambda=\lambda_1}^{\lambda_r}
p^{(\lambda)}](\frac{dA_{\tau\lambda_1\dots
\lambda_r}}{ d\xi})^2=0\quad\alpha= \tau ,\lambda_i\, .
\end{equation}
\end {itemize} We take as anz\"atze the solutions of the extremal brane  
problem but with $H_A$ an  unknown function of $H(\xi)$. Namely we pose
\begin{eqnarray}
\label{xx} A_{\tau\lambda_1\dots
\lambda_r}&=&\epsilon_{\tau
\lambda_1\dots\lambda_r}[\frac{2(D-2)}{\Delta}]^{1/2}H^{-1}(\xi)\, ,\\
\label{yy}
  p^{(\tau)}=p^{(\lambda_i)}= - \frac{D-r-3}{\Delta}  \ln H(\xi)~   &;& ~
p^{(\alpha)}=\frac{r+1}{\Delta} \ln H(\xi)\quad\alpha\neq \tau   ,\lambda_i\,
.\\
\label{zz}
\phi &=& \frac{D-2}{\Delta}\varepsilon a \ln H(\xi) \, .
\end{eqnarray} From these equations and from Eq.(\ref{delta}) we see   that
$\varepsilon a
\phi -2p^{(\tau)}-2\sum_{\lambda=\lambda_1}^{\lambda_r}   p^{(\lambda)}=2\ln
H(\xi)$.  It then follows that  the  equation of motion for $A$,  
Eq.(\ref{potential}), reduces to, using Eq.(\ref{xx}),
\begin{equation}
\label{basic}
\frac{d^2H(\xi)}{ d\xi^2}=0\, .
\end{equation} Given this result, it is straightforward to verify that the 
anz\"atze   Eqs.(\ref{xx}), (\ref{yy}) and (\ref{zz})  satisfy the
dilaton and the vielbein equations of motions. The lapse constraint takes the
form
\begin{equation}
\label{xiextremal}
\sum_{\alpha=1}^D ( d p^{(\alpha)})^2 -\frac{1}{2}(\sum_{\alpha=1}^D
dp^{(\alpha)})^2 +
\frac{1}{2} (d\phi)^2- \frac{D-2}{\Delta}(d\ln H)^2=0\, .
\end{equation} where the differentials are taken in $\xi$-space. It has  
therefore exactly the same form  in
$\xi$-space as  Eq.(\ref{extremal}) has in space-time.  The relations  
Eqs.(\ref{xx}), (\ref{yy}) and (\ref{zz}), together with Eqs.(\ref{basic}) and
the   lapse constraint Eq.(\ref{xiextremal}) fully describe an exact solution
of the full
$\G$ invariant action $\cal S$ defined recursively by Eq.(\ref{full}).   We now
discuss the significance of this result.

The  Eqs.(\ref{xx}), (\ref{yy}) and (\ref{zz})  characterise completely   the
algebraic structure of the extremal brane solution but do not yield its  
harmonic character in space-time. As the functions $A_{\tau\lambda_1\dots
\lambda_r}(\xi),\, p^{(\tau)}(\xi),\,  p^{(\lambda_i)}(\xi) $ and $  
\phi(\xi)$ were interpreted in the action $\cal S$ as functions at a fixed
space-time   point of the independent variable
$\xi$, this is a consistent result.  The solution $H = a+ b\xi$ of  
Eq.($\ref{basic}$) would then describe a motion in the space of solutions, for
instance of   branes with different charges.  However the fact that we have 
exact solutions of   the action
$\cal S$ with the correct algebraic structure of the extremal branes,   means
that these solutions  are only  indirectly related to the corresponding    
space-time solution. One expects that the information contained in this
solution,   which is of course  not contained in a trivial  constant space-time
solution of the   Einstein equation, is the required information to build
coupled equations to   higher space-time derivatives encoded in  higher level
representations, which   would then be directly related to space-time solutions.

An indication that this is indeed the case is found by exploring   the
decomposition of the adjoint representation of $E_8^{+++}$ into  
representations of
$A_{10}$. There the fields at level 1,2,3 are the 3-form  
potential at level 1, the 6-form magnetic potential at level 2 and the dual
graviton at level   3
\cite{damourhn02,west02}. The corresponding step operators are $R^{abc}$,
$R^{abcdef}$ and the dual gravitons $R^{abcdefgh,k}$. They belong to the
representations\footnote{Here we follow the usual convention. The   Dynkin
labels of the $A_{10}$ representations are labelled  from right to left when  
compared with the labelling of the Dynkin diagram of Fig.1. For instance the  
last label on the right refers to the fundamental weight associated with the
`time' root   labelled 1 in Fig.1.}
\begin{equation}
\label{rep1} (0,0,1,0,0,0,0,0,0,0)\qquad (0,0,0,0,0,1,0,0,0,0)\qquad  
(1,0,0,0,0,0,0,1,0,0)\, .
\end{equation} Arbitrarily high number $k$ of field derivatives could
tentatively be described by the   group parameters (or $\xi$-fields) of the
representations
\begin{equation}
\label{rep2} (0,0,1,0,0,0,0,0,0,k)\qquad (0,0,0,0,0,1,0,0,0,k)\qquad  
(1,0,0,0,0,0,0,1,0,k)\, .
\end{equation}
  One may verify that these indeed occur respectively at levels $7k+1, 7k+2,
7k+3$, although we have not proven, for $k>2$, that their outer multiplicity
does not vanish. This is however unlikely to happen as the
corresponding roots  are imaginary with squared length
diverging as $k\to\infty$, and one expect outer multiplicities,
which are already large\footnote{For  k=2, the outer multiplicities for the
three towers are  respectively 3290, 8567 and 36067 !  \cite{fisc}.} for $k=2$ to grow rapidly with $k$.
These towers are to some extent   analogous to
the towers which appear in
$E_8^{++}$ modulo 3
\cite{damourhn02}, but the representations modulo 7 of $E_8^{+++}$ may be highly
degenerate. Hence the
identification would require a selection rule to isolate the   relevant
representations.

Such a selection rule can be imposed by taking, for instance, the  {\em non
degenerate} (i.e. of outer multiplicity one) representation
\begin{equation}
\label{momentum} P_a=(0,0,0,0,0,0,0,0,0,1)
\end{equation}
  which appears at level 7. The negative root
$-\alpha=\lambda_1$ of $E_8^{+++}$ corresponding to a highest weight  
$P^1$ in
$A_{10}$ is given by (see Table 2 of \cite{fisch})
\begin{equation}
\label{high}
\alpha=\alpha_1+3\alpha_2+5\alpha_3+7\alpha_4+9\alpha_5+11\alpha_6+ 13\alpha_7+
15\alpha_8+10\alpha_9+5\alpha_{10}+ 7\alpha_{11}\, .
\end{equation} Labelling by
$\alpha^1_k,\alpha^2_k,\alpha^3_k$ the roots of
$E_{11}$ corresponding to the highest weights of the representations
Eq.(\ref{rep2}),  one verifies the additivity relations given in the   last
column of Table~II. These show that the generators of the representations
Eq.(\ref{rep2})  can be obtained by  multicommutators\footnote{We have however
 been unable to disprove the accidental  vanishing of some relevant
commutator.} of
$P_a$ with 
$R^{abc}$,
$R^{abcdef}$ and $R^{abcdefgh,k}$.  Thus the
generators of the towers selected in this way could be identified with the
`derivative' representations, and
$ P_a$ may be related to a generator of space-time   translations.  
One then expects the functions
  $H(\xi)$ entering the above BPS solutions to be promoted to space-time
harmonic functions when coupling with higher   levels are taken into account,
at least for $E_8^{+++}$ solutions.  From the   generality of the
correspondence between space-time and $\xi$-space appearing in the exact
solutions, we may hope for a related mechanism to be operative in all $\G$,
although it might be in general not identical to the one suggested here.
Indeed, the existence of an operator of the type displayed in
Eq.(\ref{momentum}) cannot be present\footnote{This was pointed out to us by
Axel Kleinschmidt. The periodicity argument based on the observation that
representations at any level
$l$ occur in the direct product of representations of level one implies for
$E^{+++}$ that representations with Dynkin labels Eq.(\ref{momentum}) may occur
at level
$l$ when
$10= 3l ~ mod~  11$ (or  dual ones when $1= 3l~ mod~ 11)$. Indeed one does get
such representations at level 7, 18, ... (4, 15, ...). (The representations at
level 4 and 7 have outer multiplicity one while the outer multiplicities at
level 18 and 15 are respectively 15765 and 824 \cite{fisc}.) For some $\G$, in
particular for $D_{24}^{+++}$, there are no such solutions to the periodicity
constraint \cite{klein}.} in all
$\G$. Note also that there appears to be no direct relation between
$P_a$, which is a {\em generator} of $\G$, and the tentative identification of
a momentum operator with a   weight
{\it not} contained in the adjoint representation of $\G$ \cite{west04},
as the latter would  seem to imply an extension of the symmetry   group
beyond $\G$.

\begin{center}
\begin{tabular}{||c|c|c||}
\hline Dynkin indices&level&$E_{11}$ root\\
\hline\hline
$(0,0,0,0,0,0,0,0,0,1)$&$7$&$\alpha$\\
\hline\hline
$(0,0,1,0,0,0,0,0,0,k)$&$7k+1$&$\alpha^1_k=k\alpha +\alpha^1_0 $\\
\hline
$(0,0,0,0,0,1,0,0,0,k)$&$7k+2$&$\alpha^2_k=k\alpha +\alpha^2_0$\\
\hline
$(1,0,0,0,0,0,0,1,0,k)$&$7k+3$&$\alpha^3_k=k\alpha +\alpha^3_0$\\
\hline
\end{tabular}
\vskip .3cm {\small   Table II : Infinite towers of `derivative'  
representations.}
\end{center} A more detailed account of these results and of their
implications for the generation of space-time is   differed to a separate
publication \cite{ehtp}.

\subsection{Group theory of extremal branes, KK-waves and monopoles}

The action $\cal S$ is invariant under non-linear, field dependent, $\G$
transformations. Hence one can generate new solutions of $\cal S$ from   the
extremal brane solution. In this section we shall obtain solutions   which can
be put in correspondence with  solutions of Eq.(\ref{oxid}), thereby testing  
the validity of $\G$ in this restricted domain and putting into evidence the
group-theoretical structure of these solutions.

We first note that the left hand side of the quadratic form  
Eq.(\ref{extremal}), or rather the corresponding quantity in $\xi$-space,
Eq.(\ref{xiextremal}), occurs at level zero in the full $\G$ invariant action
$\cal   S$. It appears in the first line of Eq.(\ref{linear}),  is equal
to the bilinear  form   of $\G$ restricted to its Cartan subalgebra, and  is
therefore invariant under the Weyl group of $\G$ \cite{ehtw}. The appearance of
a right-hand side in  Eq.(\ref{extremal}) has, as will now be  shown, a  non
trivial group-theoretical significance   related to generators, which do not
belong to this Cartan subalgebra but    emerge in the non-linear
realisation of
$\G$.

The solution Eqs.(\ref{xx}), (\ref{yy}) and (\ref{zz}) satisfy, in  
$\xi$-space, the relation Eq.(\ref{embedding}). This relation define an
embedding of a   subgroup
${\cal G}^{p+1}$ of $\G$ acting on the $p$  compact space dimensions    in
which the branes live and on the time dimension \cite{ehw}.  We shall consider
the subgroup  
${\cal G}^p$ of
${\cal G}^{p+1}$ which acts on the space dimensions only and we take  
$p\le D-4$ so that
${\cal G}^{p+1}$ is a Lie group. This group is conjugate by a Weyl   reflection
in  $\G$ of the group ${\cal G}^{\prime\, p+1}$ obtained by deleting the first  
$D-p-1$ nodes of the gravity line \cite{ehw} and hence ${\cal G}^p$ is conjugate
to its subgroup  
${\cal G}^{\prime\, p}$ characterising the usual dimensional reduction of  
Eq.(\ref{oxid}) to $D-p$ dimensions.

  We shall consider  the  transformations mapping {\em one}  root to  
another   root, thereby  generating solutions of the same `family' as the
extremal   solution just described.  These transformations  include the  Weyl
group $W({\G}) $
  of $\G$. We shall   examine   Weyl transforms of the extremal brane   solution
characterised by one positive step operator which  send the positive   root
into a positive root. Such transformations leave invariant not only
$\cal S$ but also preserves their quadratic truncation   Eq.(\ref{linear}).
Hence Eq.(\ref{xiextremal})  is invariant under the  Weyl group of ${\cal
G}^{p+1}$. The restriction   to 
the Weyl group of ${\cal G}^p$ selects  transformed fields with   
one time index.

Thus $W({\cal G}^p) $ leaves invariant   the quadratic form
\begin{equation}
\label{lapsegen}
\sum_{a=1}^D (dp^{(a)})^2+\frac{1}{2} [-(\sum_{i=1}^D dp^{(a)})^2  
+\sum_{u=1}^q (d\phi^u)^2 +(e^{~\mu}_b d e_\mu^{~a}d e_{a\nu}
  e^{\nu b})^{(1)}+\frac{1}{r!s!}\sum_A \,  [dA]_{a_1\dots a_r}^{~   b_1\dots
b_s}\, [dA]^{~ a_1\dots a_r}_{ b_1\dots b_s} ] ,
\end{equation} and the  embedding relation Eq.(\ref{embedding}) in  
$\xi$-space. It acts on $A$-fields, or non-diagonal vielbeins, containing one
time index.   The sum of the first three terms is the invariant metric of $\G$
restricted to its   Cartan subgroup. Together with the embedding relation
Eq.(\ref{embedding})   they are left invariant under the Weyl group of ${\cal
G}^{p+1}$. The relevance of   this invariance for extremal branes (for all
simply laced $\G$ algebras) and for   intersecting brane configurations was
pointed out in reference \cite{ehw}. However no   group-theoretical
interpretation was provided for the appearance of the  last term in  the
 relation Eq.(\ref {xiextremal}), or in  the right hand side of
Eq.(\ref{extremal}). We see here that it arises from
the Weyl transformations of the step operators and the additional terms in
Eq.(\ref{lapsegen}) guarantee   the invariance of this relation under 
$W({\cal G}^p) $.   They allow, for all $\G$, the generation by  Weyl
transformations of new   solutions from one extremal brane solution. We stress
again that in the present   approach both electric and magnetic branes are
described `electrically'.

In Appendix C, it is shown how  well-known duality symmetries of M-theory are
interpreted in the present context, including the duality transformations of
branes, KK waves and KK monopoles (Taub-NUT spaces).   Such 
transformations are however not a privilege of M-theory and occur   in {\em
all} $\G$ invariant actions. This is exemplified below, taking for  
definiteness the action $\cal S$ for the group $E_7^{+++}$ which is related to
the   action $S$ Eq.(\ref{laE7})  of gravity coupled to a 4- and a 2- form
field strength in 9 space-time dimensions. The Dynkin diagram of  
$E_7^{+++}$ is depicted in Fig.1, which exhibits the two simple electric roots
$(10)$ and $(9)$ corresponding  respectively to the step operators
$R^{7\,8\,9}$ and $R^9$ which couple  to the electric potentials
$A_{7\,8\,9}$ and
$A_9$ (see Appendix A).

We take as input the electric extremal 2-brane ${\bf e}_{(8,9)}$ in the  
directions
$(8,9)$ associated with the 4-form field strength whose corresponding  
potential is
$A_{1\,8\,9}$ and submit it to the non trivial Weyl reflection $W_{10}$  
associated with the electric root
$(10)$ of Fig.1. We display below, both for ${\bf e}_{(8,9)}$ and its  
transform,  the moduli, i.e. the vielbein components
$p^{(a)}$ and the the dilaton value $\phi$, of the brane solution  
Eqs.(\ref{yy}) and (\ref{zz}) as a ten-dimensional vector where the last
component is the   dilaton.  We also indicate the transform of the step
operator $R^{1\,8\,9}$ under   the Weyl transformation\footnote{In Appendix C
the transformations of the step   operators have been explicitly computed for
$E_8^{+++}$.}. The transformed vector follows from Eq.(\ref{we7dp}).     We
obtain

\begin{eqnarray}
\label {e89} (-4, 3,3,3,3,3,3, -4,-4;2\sqrt 7)\, \frac{\ln H(\xi)}{14}&   {\bf
e}_{(8,9)} &R^{1\,8\,9}\\ &\qquad\downarrow W_{10}\quad&\nonumber\\
\label{k17} (-7, 0,0,0,0,0,7, 0,0;0)\, \frac{\ln H(\xi)}{14}& {\bf   kk_e}_{\,
(7)} &K^1_{~7}
\end{eqnarray} The transformed brane is, according to the analysis of  
Appendix B,  a KK-wave in the direction 7, in  analogy with the double
T-duality  in   M-theory acting on the generators Eq.(\ref{kw11b}).

We now move the electric brane through Weyl reflections associated with   roots
of the gravity line to ${\bf e}_{(5 ,9)}$ and submit it to the Weyl  
reflection $W_{10}$. We now find that the brane ${\bf e}_{(5 ,9)}$ is 
invariant but   moving it to the position
${\bf e}_{(5 ,6)}$, we get
\begin{eqnarray}
\label {e56} (-4, 3,3,3,-4,-4,3,3,3;2\sqrt 7)\, \frac{\ln H(\xi)}{14}&   {\bf
e}_{(5,6)} &R^{1\,5\,6}\\ &\qquad\downarrow W_{10}\quad&\nonumber\\
\label{m56789} (-1, 6,6,6,-1,-1,-1,-1,-1;4\sqrt 7)\, \frac{\ln   H(\xi)}{14}&
{\bf m}_{(5,6,7,8,9)}  &R^{1\,5\,6\, 7\,8\,9}
\end{eqnarray} This is a magnetic 5-brane in the directions  
$(5,6,7,8,9)$ associated to the  2-form field strength~!   It is expressed in
terms   of its dual potential $A_{1\,5\, 6\,7\,8\,9}$. Submit instead ${\bf
e}_{(5 ,9)}$ to   to the Weyl reflection $W_{9}$ associated with the electric
root $(9)$ of Fig.1.   The transformed vector is now deduced from
Eq.(\ref{we79p}). The 2-brane ${\bf e}_{(5   ,9)}$  is again  invariant,
but moving it to to the position
${\bf e}_{(5 ,6)}$, we now get
\begin{eqnarray} (-4, 3,3,3,-4,-4,3,3,3;2\sqrt 7)\, \frac{\ln   H(\xi)}{14}&
{\bf e}_{(5,6)} &R^{1\,5\,6}\nonumber \\ &\qquad\downarrow  
W_{9}\quad&\nonumber\\
\label{m569} (-3, 4,4,4,-3,-3,4,4,-3;-2\sqrt 7)\, \frac{\ln   H(\xi)}{14}& {\bf
m}_{(5,6,9)}  &R^{1\,5\,6\, 9}
\end{eqnarray} This is a magnetic 3-brane in the directions $(5,6,9)$  
associated to the  4-form field strength, expressed in terms of its dual
potential  
$A_{1\,5\, 6\,9}$.  These properties are reminiscent of the double T-duality  in
M-theory   acting on the generators Eq.(\ref{r3w11a}) and (\ref{r3w11b}), but
are more involved,   due to the interplay of the 2- and 4- form field strength
in the  action   Eq.(\ref{laE7}).

Finally, let us submit the magnetic 5-brane $ {\bf m}_{(5,6,7,8,9)}$   obtained
in Eq.(\ref{m56789}) to the Weyl reflection  $W_{9}$. One obtains
\begin{eqnarray} (-1, 6,6,6,-1,-1,-1,-1,-1;4\sqrt 7)\, \frac{\ln   H(\xi)}{14}&
{\bf m}_{(5,6,7,8,9)}  &R^{1\,5\,6\, 7\,8\,9}\nonumber \\ &\qquad\downarrow
W_{9}\quad&\nonumber\\
\label{h2349} (0, 7,7,7,0,0,0,0,-7;0)\, \frac{\ln H(\xi)}{14}&  {\bf  
kk_m}_{\, (2,3,4;9)} &R^{1\,5\,6\,7\,8\,9,\,9}
\end{eqnarray} Eq.(\ref{h2349}) describes, as in M-theory under the
transformation  Eq.(\ref{r6w11b}), a purely gravitational     configuration,
namely a KK-monopole  (see Appendix B Eq.(\ref{modkkmono})) with transverse  
directions (2,3,4) and Taub-NUT direction (9) in terms of a dual gravity tensor
$h_{1\,5\,6\,7\,8\,9,\,9}$.

It is also possible, as in Appendix C,  to generate solutions not   contained,
at least explicitly, in the group ${\cal G}^p$. These are very interesting  
solutions as they may test the significance of genuine Kac-Moody extensions of
the Lie   groups. Such analysis is outside the scope of the present paper where
we test only   solutions which can straightforwardly be mapped  to space
time solutions of the effective   actions Eq.(\ref{oxid}).

The above example illustrate the analogy of M-theory duality   transformations
with similar transformations in all `M-theories' defined by all $\G$.  One  
may indeed carry the same analysis for all $\G$, using the results of Appendix
A, and exhibit for each of them the   `duality' transformations of the branes.
As in M-theory, these dualities are   symmetries in non-compact space-time. 
This is because
${\cal G}^{ p+1}$ is, as ${\cal G}^{\prime\, p+1}$, the Lie group   symmetry 
of the action Eq.(\ref{oxid}) dimensionally reduced to three dimensions (for  
$p=D-4$). They differ because while the latter reduction leaves a Lorentzian  
non-compact space-time, the former leads to a Euclidean space-time by  
compactifying time. The group ${\cal G}^p$ of transformations on $\xi$-space 
discussed   above, is thus in one to one correspondence with the group ${\cal
G}^p$ of space-time transformations when time is decompactified. In particular,
the   functions
$H(\xi)$ can thus be mapped into  harmonic functions $H(\{x^\nu\})$.   However
as pointed out in the previous section, more work is needed to relate   directly
$H(\xi)$ to
$H(\{x^\nu\})$, and solutions in
$\xi$-space to solutions in space-time for all $\G$, through translation  
operators hopefully induced by group generators, as suggested by
Eq.(\ref{momentum}).

\setcounter{equation}{0}

\section{Conclusion}

The basic concept underlying the present approach is the tentative   inclusion
of {\em local} symmetries in an infinite {\em global} symmetry. We find the
results obtained in the present work indicative of the possibility of   such an
embedding whose consequences for consistent theories of gravity and   matter
may be far reaching. They would not require an explicit   implementation of
diffeomorphism and gauge invariance which should emerge dynamically and might
lead to the inclusions of many new degrees of freedom.

We have constructed in a recursive way invariant actions    for all global
symmetries
$\G$. They are not defined in space-time but there are definite   indications
that the latter may emerges from $\G$.  Further work is necessary to  
ascertain wether or not generators of
$E_8^{+++}$ can be related to space-time translations  and if some encoding
of space-time can be generalised  for all $\G$. Exact  solutions have been
presented, which contain all the   algebraic structure of  extremal BPS
charged  brane solutions and BPS   gravitational branes, namely Kaluza-Klein
waves and KK-monopoles (Taub-NUT spaces). The   transformation properties of
these solutions put into evidence the general group-theoretical   origin of
`dualities' for all $\G$, which
 apparently do not require  an   underlying string theory.

These transformations also allow for solutions, alluded to in Section 3.3 and
Appendix C, which are not related in an obvious way to  
   Einstein's gravity. Such solutions occur when a step operator associated
with a parameter of   low level, which does have an  interpretation in terms of
a field of the    oxidised theory, is mapped under Weyl transformation
on a step operator of higher level. The significance of such solutions remains
to be unveiled. Further study should tell whether  they fit into the  
framework of general relativity as we know it, or into some generalisation of
it. Recall that  the effective action of the bosonic string and of the bosonic
sector of M-theory are   maximally oxidised theories and hence can be
extended to
$\G$ theories. One may   hope that the degrees of freedom of string theories,
possibly in the tensionless   limit, have their counterpart in the $\G$
theories. In such a perspective the present approach  may constitute a first
step towards a   consistent theory of gravity which would encompass string
theories and  allow   for different   matter content.

\section*{Acknowledgments}

We are very much indebted to Marc Henneaux  for many
fruitful discussions, and to Hermann Nicolai for illuminating conversations
during the completion of this work. We are grateful to Axel Kleinschmidt for
informative comments and to Thomas Fischbacher for providing us with a list of
the representations of
$A_{10}$ in the decomposition of $E_{11}$ up to level 19, which he produced
with an impressive mastering of the problem. One of us (F.E.) would like to
thank  the Racah Institute of the Hebrew University of Jerusalem, where part of
this work was completed, and in particular Eliezer Rabinovici, for the warm
hospitality. He expresses his gratitude to Professor Joseph Katz and to his
wife Ruthi, who contributed immensely to make his stay in Jerusalem comfortable
and fruitful. 

\noindent
This work was supported in part  by the NATO
grant PST.CLG.979008, by the ``Actions de Recherche Concert\'ees'' of the
``Direction de la Recherche Scientifique - Communaut\'e Fran\c caise de
Belgique, by a ``P\^ole d'Attraction Interuniversitaire'' (Belgium), by IISN
convention 4.4505.86 (Belgium) , by Proyectos FONDECYT 1020629, 1020832 and
7020832 (Chile),  by the European Commission RTN programme HPRN-CT00131, in
which the authors are associated to the Katholieke Universiteit te Leuven
(Belgium), by the BSF American-Israel Bi-National Science Foundation, by the
Israel Academy of Sciences and Humanities - Centers of Excellence Program and
by the German-Israel Bi-National Science Foundation.

\newpage
\appendix
\setcounter{equation}{0}
\renewcommand\theequation{\thesection.\arabic{equation}}

\section{${\cal G}^{+++}$ and Weyl reflexions}

\subsection{$A_{D-3}^{+++}$}

This case corresponds to pure gravity. Its Dynkin diagram is displayed in
Fig.1. The rank of $A_{D-3}^{+++}$  is $n=D$

The simple roots of $A_{D-3}^{+++}$ are
$E_m= K^m {}_{m+1},\ m=1,\dots ,n-1$ and $E_{n}= R^{4\ldots   n,n}$   where
$R^{a_1\dots a_{n-3},b}$ is totally anti-symmetric in its $a$ indices     and
the part antisymmetrised in all its indices  vanishes
\cite{lambertw01,west02}.   
  This generator of
$A_{D-3}^{+++}$  transforms under
$SL(D)$ according to Eq.(\ref{tensor}),
\begin{equation} [K^a{}_b, R^{c_1\dots     c_{n-3},d}]=\delta_b^{c_1}R^{a\dots
c_{n-3},d}+\dots +\delta_b^d R^{c_1\dots c_{n-3},a}\, .
\end{equation}  The Cartan generators in the Chevalley basis are given   by
\cite{ehtw}
\begin{eqnarray} H_m&=& (K^m{}_m-K^{m+1}{}_{m+1})\qquad a=1,\dots,n-1\,
\nonumber\\ H_{n}&=&-(K^1{}_1 +K^2{}_2+K^{3}{}_{3}) + K^{n}{}_{n}\, .
\end{eqnarray} The non-trivial Weyl reflexion  $W_n$, namely the one which does
not   correspond to a simple root in the gravity line  but to the simple root
$\alpha_n$ has already be discussed in \cite{ehtw}. This Weyl reflexion  
induces the following changes on the parameters
\begin{eqnarray}
\label{wna}  p^{\prime m}&=&p^m+(p^4+\ldots +p^{n-1})+2p^n\qquad m=1,2,3
\nonumber\\  p^{\prime m}&=&p^m\qquad m=4,\dots ,n-1\nonumber\\ p^{\prime
n}&=&-p^n-(p^4+\ldots +p^{n-1})\, .
\end{eqnarray}

\subsection{$B_{D-2}^{+++}$ }

This non simply laced case has already been sketched in \cite{ehtw}.   Its
Dynkin diagram is displayed in Fig.1.   The maximally oxidised theory
associated with $B_{D-2}^{+++}$  is characterised by the following   action
\begin{eqnarray}
\label{laB} S  &=& \int d^Dx \, \sqrt{-g^{(D)}}\left(R^{(D)}-
\frac{1}{2}\partial_\mu\phi\partial^\mu\phi-{1\over 2 . 3! }e^{-
l\phi}F_{\mu\nu\sigma}F^{\mu\nu\sigma}\right.\nonumber\\   &&\left.-{1\over 2.2!
}e^{- (l/2)\phi}F_{\mu\nu}F^{\mu\nu}+   C.S.\right)\, ,
\end{eqnarray} where $l$ is given by
\begin{equation}
\label{adil} l =\left({8\over D-2}\right)^{1/2}\, .
\end{equation} The rank $n$ of $B_{D-2}^{+++}$ is $n=D+1$. Thus one   must add
to the $GL(D)$ generator $K^a{}_b$ a commuting generator
$R$ associated  with the dilaton. In addition to the simple step   operators
$K^m{}_{m+1}$ one has two other  step operators $R^D$ and
$R^{5 \ldots D}$  associated to the two simple roots, one short and one   long,
which do not belong to  the gravity line. These  generators belongs to  the
$A_{D-1}$ representations $R^a$   and
$R^{a_1 \dots a_{n-5}}$  that are respectively a 1 and $n-5$ rank  
anti-symmetric tensors. From Eq.(\ref{dilaton}) they satisfy the following 
equations
\begin{equation}
\label{dilB} [R, R^a]={l\over4}R^{a}\, , ~~~ [R,   R^{a_1\ldots  
a_{n-5}}]=-{l\over 2} R^{a_1\ldots a_{n-5}}~\,  .
\end{equation} The Cartan generators of $B_{D-2}^{+++}$ in the   Chevalley
basis are given by
\begin{eqnarray}
\label{cartanb} H_m&=&(K^m{}_m-K^{m+1}{}_{m+1}) \qquad   m=1,\dots,n-2\\
\label{cartanbn-1} H_{n-1}&=&-{2\over (D-2)}(K^1{}_1+\ldots +K^{n-2}{}_{n-2}) +
{2 (D-3)\over (D-2)}K^{n-1}{}_{n-1}+l\, R\\ \label{cartanbn}
H_{n}&=&-{(D-4)\over (D-2)}(K^1{}_1+\ldots +K^4{}_4)+{2\over  
(D-2)}(K^5{}_5+\ldots K^{n-1}{}_{n-1}) - l\, R\, .
\end{eqnarray} There are two non-trivial Weyl reflexions $W_{n-1}$ and  
$W_n$
  corresponding  respectively to the simple roots $\alpha_{n-1}$ and
$\alpha_n$.

\bigskip\bigskip\bigskip
\noindent
\underline{\large  $W_{n-1}$}

\bigskip\bigskip
\noindent The  Weyl reflexion induces the following changes on the   generators
\begin{eqnarray}
\label{wn-1g}  K^m{}_m &\to& K^m{}_m\qquad a=1,\ldots ,n-2 \,\nonumber\\
K^{n-1}{}_{n-1}&\to&  K^{n-1}{}_{n-1} -H_{n-1} \nonumber\\ R &\to& R   -{2\over
l(D-2)}H_{n-1}\, ,
\end{eqnarray} which yield the parameter  transformations 
\begin{eqnarray}
\label{wn-1p}
  p^{\prime m}&=&p^m +{2\over (D-2)}p^{n-1}+{l\over 2\, (D-2)}\phi   
\qquad m=1,\dots  n-2\nonumber\\ p^{\prime n-1}&=&{(4-D)  
\over(D-2)}p^{n-1}-{l\, (D-3)\over 2\, (D-2)}\phi \nonumber\\
\phi^{\prime}&=&{(D-4)\over   (D-2)}\phi -l\, p^{n-1}\, .
\end{eqnarray}

\bigskip
\noindent {\underline{\large  $W_n$}}

\bigskip
\noindent The  Weyl reflexion induces the following changes on the   generators
\begin{eqnarray}
\label{wng}  K^m{}_m &\to& K^m{}_m\qquad m=1,\ldots ,4 \,\nonumber\\ K^m{}_m 
&\to& K^m{}_m -H_n \qquad m=5,\ldots ,n-1 \,\nonumber\\ R    &\to& R +{l\over
2}H_n\, ,
\end{eqnarray} which yield the parameter  transformations 
\begin{eqnarray}
\label{wnp} p^{\prime m}&=&p^m+{D-4\over D-2}(p^{5}+\dots +p^{n-1})     -{l\,
(D-4)\over 2\, (D-2)}\phi\qquad m=1,\dots ,4 \nonumber\\ p^{\prime m}&=& p^m
-{2\over D-2}(p^{5}+\ldots +p^{n-1})   +{l\over (D-2)}\phi\qquad   m=5,\ldots
,n-1\qquad \nonumber \\ \phi^{\prime}&=&\phi+l\, (p^{5}+\ldots   +p^{n-1})
-{4\over (D-2)}\phi\, .
\end{eqnarray}

\subsection{$C_{q+1}^{+++}$}

The Dynkin diagram characterising this non-simply laced theory is given   in
Fig.1. The corresponding  maximally oxidised theory is defined in four
dimensions   and  is characterised by $q$ different dilatons. The action is  
\cite{cremmerjlp99}
\begin{eqnarray}
\label{laC} S &=& \int d^4x \, \sqrt{-g^{(4)}}\left(R^{(4)}-
\frac{1}{2} \sum_{u=1}^{q} \partial_\mu\phi^u\partial^\mu\phi^u   -{1\over 2  }
\sum_{\alpha}  e^{\sigma_{\alpha} . \phi} F^{\alpha }_{\mu}F^{\alpha
\mu}\right.\nonumber\\ &&\left. -{1\over 2.2! } \sum_{i=1}^q e^{e_i .
\phi}F^{i}_{\mu\nu} F^{i\mu\nu} \right)\, ,
\end{eqnarray} where $\sigma_{\alpha} =\{ 2e_i,e_j  \pm e_i \,  {\rm   with} \,
j>i \}$ and the $e_i$  are $q$-dimensional vectors $(0,\dots ,1, \dots 0)$
with   the $1$ at the
$i^{\rm th}$ position.  For each dilaton $\phi^u$ with $ u=1 \dots q$   one
introduces a  generator $R_u$ commuting with the $K^a{}_b$.

The simple roots which do not belong to the gravity line are of two   kinds.
First there are the   electric roots of the one-form field strengths  
$F^\alpha_1$ with $\alpha$ such that
$\sigma_{\alpha}= e_{i+1}-e_i $ with $i=1 \dots q-1$ and
$\sigma_{\alpha}=2 e_1$. We associate with these roots respectively  
$q-1$ generators denoted $S_j$ with $j=1, \dots ,q-1$ and a generator $S_q$.  
They are in a scalar representation of $A_3$. Second there is the magnetic
root   associated to the two-form field strength
$F_2^q$ with dilaton coupling $e_q$. We associate with this simple root   a 
step operator $R^4$ which belong to a vectorial representation $R^a$ of  
$A_3$.

To summarise we have the following simple step operators for  
$C_{q+1}^{+++}$
\begin{eqnarray} E_m &=& K^m{}_{m+1}\qquad m=1,2,3\nonumber\\ E_4 &=&
R^4\qquad,\qquad E_{4+j} = S_j \qquad j=1, \dots ,q\, .
\end{eqnarray} We  deduce from the lagrangian Eq.(\ref{laC}),  knowing   the
electric or magnetic nature of the simple roots and Eq.(\ref{dilaton}),  
  the following relations for $u=1, \dots ,q$
\begin{eqnarray}
\label{comC}  [R_u, R^{a}]& = &{1 \over 2} \delta_{q,u} \,   R^{a}\nonumber\\
\strut [R_u, S_j ] &=& -{1 \over 2} ( \delta_{u,q+1-j}-\delta_{u,q-j} )  
\, S_j
\qquad j=1,\dots ,q-1 \nonumber\\ \strut [R_u,S_q] &=& -\delta_{u,1} \,   S_q\,
.
\end{eqnarray} From the commutators Eq.(\ref{comC}) we can  determine the
Cartan generators in the Chevalley basis in terms of those the   K-basis. We get
\begin{eqnarray}
\label{cartanc} H_m&=&(K^m{}_m-K^{m+1}{}_{m+1})\qquad~ m=1,\dots,3\\
\label{cartanc4} H_{4}&=&-(K^1{}_1+K^{2}{}_{2}+K^{3}{}_{3} )+K^{4}{}_{4} +2 \,
R_q\\
\label{cartancn} H_{4+m}&=&-2\, R_{q+1-m} +2\, R_{q-m} \qquad m=1, \dots ,q-1\\
\label{cartancq} H_{4+q}&=& -2 \,  R_1\,  .
\end{eqnarray} We are now in position to describe the non-trivial Weyl  
reflexions

\bigskip\bigskip\bigskip
\noindent {\underline{\large  $W_4$}}

\bigskip\bigskip
\noindent The  Weyl reflexion corresponding to the simple root
$\alpha_4$ induces the following changes on the generators
\begin{eqnarray}
\label{wcg}  K^m{}_m &\to& K^m{}_m\qquad a=1,\ldots ,3 \,\nonumber\\    K^4{}_4
&\to& K^4{}_4 -H_4 \nonumber\\  R_q &\to& R_q -{1\over 2}H_4\, ,
\end{eqnarray} which yield the parameter  transformations 
\begin{eqnarray}
\label{wcp} p^{\prime m}&=&p^m+p^{4}+ {1 \over 2} \phi^{q}
\qquad m=1,\dots ,3 \nonumber\\ p^{\prime 4}&=& -{1\over   2}\phi^q\nonumber
\\
\phi^{\prime q} &=& -2 p^4 \nonumber\\\phi^{\prime u}&=& \phi^u \qquad   u=1,
\dots q-1\, .
\end{eqnarray}

\bigskip
\noindent
\underline{\large  $W_{4+j}$}, $\qquad (j=1, \dots ,q-1)$

\bigskip
\noindent Using Eqs.(\ref{cartanc})-(\ref{cartancq}) one gets the Weyl
reflexions $W_{4+j}$ for $j=1, \dots q-1$. They induce the following
transformations on the parameters
\begin{eqnarray}
\label{weylcn} && p^{\prime m}=p^m \qquad m=1,\dots ,4 \nonumber\\
&&\phi^{\prime q+1-j} =  \phi^{q-j} \quad,\quad
\phi^{\prime q-j} =  \phi^{q+1-j} \, .
\end{eqnarray} The others $\phi^u$ are left invariant. These   transformations
interchange  two neighbouring dilatons.

\bigskip\bigskip
\noindent
\underline{\large  $W_{4+q}$}

\bigskip
\noindent Using Eqs.(\ref{cartanc})-(\ref{cartancq}) one gets the Weyl
reflexions $W_{4+q}$. They induce the following transformation of the  
parameters
\begin{eqnarray}
\label{weylcf} p^{\prime m}&=&p^m \qquad m=1,\dots ,4 \nonumber\\
\phi^{\prime u} &=&  \phi^{u}\qquad u=2, \dots, q \nonumber\\  
\phi^{\prime 1} &= &- \phi^{1}\, .
\end{eqnarray} This is an $S$-duality like transformation on the first  
dilaton.

\subsection{$D_{D-2}^{+++}$}

The rank $n$ is $n=D+1$. There is thus a dilaton and its associated generator
$R$. The Dynkin diagram characterising this theory is given   in Fig.1. The
corresponding  maximally oxidised theory is defined in D dimensions   and for
$D=26$ it is the low-energy effective action of the bosonic string   (without
the tachyon). The action is
\begin{equation}
     \label{lad}
    S =  \int d^Dx \, \sqrt{-g^D}\left(R^D-
\frac{1}{2}\partial_\mu\phi\partial^\mu\phi-{1\over 2 . 3!
}e^{-l\phi}F_{\mu\nu\sigma}F^{\mu\nu\sigma}\right)\, ,
\end{equation} with $l$ given by Eq.(\ref{adil}). $D_{D-2}^{+++}$ has   already
been discussed in the literature \cite{west01,west05,ehtw} and in  particular
the non-trivial Weyl reflexions have been presented in
\cite{ehtw}. We list here the results for sake of completness.

The Cartan generators are given by \cite{west01} (see \cite{ehtw} for the
 normalisation used here)
\begin{eqnarray} H_m&=&K^m{}_m-K^{m+1}{}_{m+1}\qquad m=1,\ldots,n-2\nonumber\\
H_{n-1}&=&-{2\over(D-2)}(K^1{}_1+\ldots +K^{n-3}{}_{n-3}){(D-4)\over
(D-2)}(K^{n-2}{}_{n-2}+ K^{n-1}{}_{n-1})+l\, R~~~~~~~\nonumber\\
H_{n}&=&-{(D-4)\over(D-2)}(K^1{}_1+\ldots +K^{4}{}_{4})+{2\over
(D-2)}(K^{5}{}_{5}+\ldots + K^{n-1}{}_{n-1})-l\, R\, .
\end{eqnarray} There are two non-trivial Weyl reflexion $W_{n-1}$ and $W_{n}$   
corresponding respectively to the simple roots $\alpha_{n-1}$ and
$\alpha_n$. They  induce the parameter  transformations 

\bigskip\bigskip
\noindent
\underline{\large  $W_{n-1}$}

\bigskip
\noindent
\begin{eqnarray}
\label{Wedn-1} p^{\prime m}&=&p^m+{2\over (D-2)}(p^{n-2}+p^{n-1})   +{l\over
(D-2)}\phi\qquad m=1,\dots ,n-3 \nonumber\\ p^{\prime m}&=&p^m -{(D-4)\over
(D-2 )}(p^{n-2}+p^{n-1}) -{l\,(D-4)\over 2\,   (D-2)}\phi\qquad
m=n-2,n-1\nonumber\\ \phi^{\prime}&=& {(D-6)\over
(D-2)}\phi-l(p^{n-2}+p^{n-1})\, .
\end{eqnarray}

\bigskip\bigskip
\noindent
\underline{\large  $W_{n}$}

\bigskip
\noindent
\begin{eqnarray}
\label{Wedn} p^{\prime m}&=&p^m+{D-4\over D-2}(p^{5}+\dots +p^{n-1})     -{l\,
(D-4)\over 2\, (D-2)}\phi\qquad m=1,\dots ,4 \nonumber\\ p^{\prime   m}&=&p^m
-{2\over D-2}(p^{5}+\ldots +p^{n-1})   +{l\over (D-2)}\phi\qquad   m=5,\ldots
,n-1\nonumber\\\phi^{\prime}&=&  {(D-6) \over (D-2)}\phi+ l\,   (p^{5}+\ldots
+p^{n-1})\, .
\end{eqnarray}

\subsection{$E_6^{+++}$}

The Dynkin diagram characterising this theory is given in Fig.1. The
corresponding  maximally oxidised theory is defined in eight dimensions  
  and  is characterised by one dilaton, a four-form   and a one-form field
strength. The action is given by \cite{cremmerjlp99}
\begin{eqnarray}
\label{lae6} S &=&  \int d^8x \, \sqrt{-g^{(8)}}\left(R^{(8)}-
\frac{1}{2}\partial_\mu\phi\partial^\mu\phi-{1\over 2 . 4! }e^{-
\phi}F_{\mu\nu\sigma\rho}F^{\mu\nu\sigma\rho}\right.\nonumber\\
&&\left.-{1\over 2 }e^{2\phi}F_{\mu}F^{\mu}+   C.S.\right)\, .
\end{eqnarray} Once again there is a generator $R$ associated to the   dilaton
and commuting with the $ K^a{}_b$. In addition to  the simple step   operators
of the gravity line one has two other step operators
  $R^{\,6\,7\,8}$ and $S$ corresponding respectively to the electric   simple
root of the four-form field strength and to electric simple root of the  
one-form field strength. The $R^{\,6\,7\,8}$ generator belongs to the $A_7$  
representation  which is a third rank antisymmetric tensor $R^{abc}$ obeying
thus    Eq.(\ref{root}). The remaining  generator $S$ is a $A_7$-scalar.
Considering the dilaton   couplings in Eq.(\ref{lae6}) and using
Eq.(\ref{dilaton}), one gets the following   commutators
\begin{equation}
\label{stepe6c} [R,R^{abc}]={1\over 2} \, R^{abc}\, , \qquad [R,S]=-S\,   .
\end{equation} Using Eq.(\ref{stepe6c}) one can then express the Cartan  
generator in  terms of the $(K,R)$-basis. One gets
\begin{eqnarray}
\label{cartane6} H_m&=&(K^m{}_m-K^{m+1}{}_{m+1}) \qquad   m=1,\dots,7\\
H_{8}&=&-{1 \over 2}(K^1{}_1+\ldots +K^{5}{}_{5})  +{1 \over
2}(K^{6}{}_{6}+K^{7}{}_{7}+K^{8}{}_{8})    +\, R\label{cartane68}\\
\label{cartane69} H_{9}&=&- 2\, R\, .
\end{eqnarray} There are two non-trivial Weyl reflexions  $W_8$ and  
$W_9$
  corresponding  respectively to the simple roots $\alpha_8$ and
$\alpha_9$.

\bigskip\bigskip\bigskip
\noindent
\underline{\large  $W_{8}$}

\bigskip\bigskip
\noindent The  Weyl reflexion induces the following changes on the   generators
\begin{eqnarray}
\label{we68g}  K^m{}_m &\to& K^m{}_m\qquad a=1,\ldots ,5 \,\nonumber\\ K^m{}_m
&\to& K^m{}_m -H_8 \qquad a=6 \ldots 8 \nonumber\\  R &\to& R -{1\over 2}H_8\, ,
\end{eqnarray} which yield the parameter  transformations
\begin{eqnarray}
\label{we68p} p^{\prime m}&=&p^m+{1\over 2}(p^{6}+p^{7}+p^{8})     +{1\over
4}\phi\qquad m=1,\dots ,5 \nonumber\\ p^{\prime m}&=&p^m -{1\over
2}(p^{6}+p^{7} +p^{8})   -{1\over 4}\phi\qquad m=6,\ldots ,8 \nonumber  
\\
\phi^{\prime}&=&{1\over 2} \phi - (p^{6}+p^{7} +p^{8})\, .
\end{eqnarray}

\bigskip
\noindent
\underline{\large  $W_{9}$}

\bigskip
\noindent This Weyl reflexion gives 
\begin{eqnarray}
\label{we69p} p^{\prime m}&=&p^m\qquad m=1,\dots ,8 \nonumber\\
\phi^{\prime}&=&-\phi\, ,
\end{eqnarray}  which is an $S$-duality like transformation.

\subsection{$E_7^{+++}$}

The Dynkin diagram characterising this theory is given in Fig.1. The
corresponding  maximally oxidised theory is defined in  nine dimensions   and 
is characterised by one dilaton, a four-form field strength  and a   two-form
field strength. The action is given by \cite{cremmerjlp99}
\begin{eqnarray}
\label{laE7} S &=&  \int d^9x \, \, \sqrt{-g^{(9)}}\left(R^{(9)}-
\frac{1}{2}\partial_\mu\phi\partial^\mu\phi-{1\over 2 . 4! }e^{l
\phi}F_{\mu\nu\sigma\rho}F^{\mu\nu\sigma\rho}\right.\nonumber\\
&&\left.-{1\over 2.2! }e^{-2l\phi}F_{\mu\nu}F^{\mu\nu}+   C.S.\right)\,   ,
\end{eqnarray} with
\begin{equation}
\label{lvalue} l={2 \over \sqrt{7}}\, .
\end{equation} One has a  generator $R$ associated with the dilaton and
commuting with the $ K^a{}_b$. In addition to  the simple step   operators of
the gravity line one has two other step operators
  $R^{\,7\,8\,9}$ and $R^9$ corresponding respectively to the electric   simple
root of the four-form field strength and to electric simple root of the  
two-form field strength. The $R^{\,7\,8\,9}$ generator belongs to the $A_8$  
representation  which is a third rank antisymmetric tensor $R^{abc}$ and the
remaining    generator $R^9$ belongs to  a $A_8$-vector $R^a$. Considering the
dilaton couplings in Eq.(\ref{laE7}) and using Eq.(\ref{dilaton}), one gets the
following   commutators
\begin{equation}
\label{stepe7c} [R,R^{abc}]= -{l \over 2} \, R^{abc}\, , \qquad   [R,R^a]=l \,
R^a\, .
\end{equation} Using Eq.(\ref{stepe7c}) one can then express the Cartan  
generator in  terms of the $(K,R)$-basis. One gets
\begin{eqnarray}
\label{cartane7} H_m&=&(K^m{}_m-K^{m+1}{}_{m+1}) \qquad
m=1,\dots,8\\\label{cartane79} H_{9}&=&-{1 \over 7}(K^1{}_1+\ldots  
+K^{8}{}_{8}) +{6 \over   7}K^{9}{}_{9}    +2l\, R\\ \label{cartane7d}
H_{10}&=&-{3  
\over 7}(K^1{}_1+\ldots +K^{6}{}_{6})  +{4 \over
7}(K^{7}{}_{7}+K^{8}{}_{8}+K^{9}{}_{9})    - l\, R\,  .
\end{eqnarray} There are two non-trivial Weyl reflexions $W_9$ and  
$W_{10}$
  corresponding  respectively to the simple roots $\alpha_9$ and
$\alpha_{10}$.

\bigskip
\noindent
\underline{\large  $W_9$}

\bigskip
\noindent The  Weyl reflexion induces the following changes on the   generators
\begin{eqnarray}
\label{we79g}  K^m{}_m &\to& K^m{}_m\qquad a=1,\ldots ,8 \,\nonumber\\
K^9{}_9&\to& K^9{}_9 -H_9\nonumber\\ R &\to& R -l\, H_9\, ,
\end{eqnarray} which yield the parameter  transformations
\begin{eqnarray}
\label{we79p} p^{\prime m}&=&p^m+{1\over 7}p^{9}   +{l\over 7}\phi\qquad
m=1,\dots ,8 \nonumber\\p^{\prime 9}&=&{1\over 7}p^9-{6l\over 7}\phi\nonumber
\\ \phi^{\prime}&=&-{1\over 7} \phi -2l\, p^{9}\, .
\end{eqnarray}

\bigskip
\noindent
\underline{\large  $W_{10}$}

\bigskip
\noindent This Weyl reflexion gives 
\begin{eqnarray}
\label{we7dp} p^{\prime m}&=&p^m+{3\over 7}(p^{7}+p^{8}+p^{9}   -{l\over 2}\phi)
\qquad m=1,\dots ,6 \nonumber\\ p^{\prime m}&=&p^m-{4\over
7}(p^{7}+p^{8}+p^{9}   -{l\over 2}\phi)\qquad m=7,\dots ,9 \nonumber\\
\phi^{\prime}&=&{5\over 7} \phi + l\, (p^{7}+p^{8}+p^{9})\, .
\end{eqnarray}

\subsection{$E_8^{+++}$}

The Dynkin diagram characterising this theory is given in Fig.1. The
corresponding  maximally oxidised theory is the bosonic sector of eleven
dimensional supergravity whose action is
\begin{equation}
     \label{lae8}
    S= \int d^{11}x \, \sqrt{-g^{(11)}}\left(R^{(11)}- {1\over 2  . 4!
}F_{\mu\nu\sigma\tau}F^{\mu\nu\sigma\tau}+ CS \right)\, .
\end{equation} $E_8^{+++}$ has already been extensively studied in the  
literature
\cite{west01,west02, ehtw} and in  particular the non-trivial Weyl reflexions
have been  presented in
\cite{ehtw}. Again we list them here for sake of completness.

The Cartan generator are given by \cite{west01}
\begin{eqnarray}
\label{aa} H_m&=& (K^m{}_m-K^{m+1}{}_{m+1})\qquad  m=1,\dots,10  
\nonumber\\
\label{el} H_{11}&=&-{1\over 3}(K^1{}_1+\ldots +K^8{}_8) +{2\over    
3}(K^9{}_9+ K^{10}{}_{10}+K^{11}{}_{11})\,.
\end{eqnarray}

The non-trivial Weyl transformation $W_{11}$ corresponding to the   simple root
$\alpha_{11}$ is
\begin{eqnarray} &&K^{\prime a}{}_a=K^a{}_a\qquad a=1,\ldots ,8
\nonumber\\\label{alpha11} &&K^{\prime a}{}_{a}=K^a{}_a -H_{11} \qquad
a=9,10,11\, .
\end{eqnarray} yielding
\begin{eqnarray}
\label{Wee8} p^{\prime a}&=&p^a +\frac{1}{3}(p^9+p^{10}+p^{11})
\qquad a=1,\dots, 8 \nonumber\\ p^{\prime a}&=&p^a
-\frac{2}{3}(p^9+p^{10}+p^{11})
\qquad a=9,10,11\, .
\end{eqnarray}

\subsection{$F_4^{+++}$}

The Dynkin diagram characterising this non simply laced theory is given   in
Fig.1. The corresponding  maximally oxidised theory is defined in  six
dimensions   and  is characterised by one dilaton, a one-form field strength,
  two two-form field strengths and two three-form field strengths.  The   action
is given by \cite{cremmerjlp99}
\begin{eqnarray}
\label{laf4} S &=& \int d^6x \, \sqrt{-g^{(6)}}\left(R^{(6)}-
\frac{1}{2}\partial_\mu\phi\partial^\mu\phi -{1\over 2 . 3!
}e^{-l\phi}F^{(1)}_{\mu\nu\sigma}F^{(1)\mu\nu\sigma}\right. -{1\over 2   . 3!
}F^{(2)}_{\mu\nu\sigma}F^{(2)\mu\nu\sigma}
\nonumber\\ &&\left. -{1\over 4}e^{-{l \over  
2}\phi}F^{(1)}_{\mu\nu}F^{(1)\mu\nu} - {1\over 4}e^{{l\over 2}\phi}
F^{(2)}_{\mu\nu}F^{(2)\mu\nu} -{1\over 2 }e^{l\phi}F_{\mu}F^{\mu}
+C.S.\right)\, .
\end{eqnarray} with
\begin{equation}
\label{lvalue2} l=\sqrt{2}\, .
\end{equation} One has a  generator $R$ associated with the dilaton and
commuting with the $ K^a{}_b$. In addition to  the simple step   operators of
the gravity line one has two other step operators
  $R^6$ and $S$ corresponding respectively to the electric simple  root   of the
two-form field strength $F_2^{(1)}$ and to electric simple root of  the  
one-form field strength. The $R^6$ generator belongs to the vectorial $A_5$   
representation and the remaining  generator $S$ is a   $A_5$ scalar. 
Considering the   dilaton couplings in Eq.(\ref{laf4}) and using
Eq.(\ref{dilaton}), one gets the   following commutators
\begin{equation}
\label{stepef4} [R,R^a]= +{l \over 4} \, R^a\, , \qquad [R,S]=-{l \over   2} \,
S\, .
\end{equation} Using Eq.(\ref{stepef4}) one can  express the Cartan   generator
in  terms of the $(K,R)$-basis. One gets
\begin{eqnarray}
\label{cartanf} H_m&=&(K^m{}_m-K^{m+1}{}_{m+1}) \qquad
m=1,\dots,5\\\label{cartanf6} H_{6}&=&-{1 \over 2}(K^1{}_1+\ldots  
+K^{5}{}_{5}) +{3 \over   2}K^{6}{}_{6}    + {2\over l}\, R\\
\label{cartanf7} H_{7}&=&- {4 \over l}\, R\, .
\end{eqnarray} There are two non-trivial Weyl reflexions  $W_6$ and  
$W_7$
  corresponding  respectively to the simple roots $\alpha_6$ and
$\alpha_7$.

\bigskip
\noindent
\underline{\large  $W_6$}

\bigskip
\noindent The  Weyl reflexion induces the following changes on the   generators
\begin{eqnarray}
\label{wef6g}  K^m{}_m &\to &K^m{}_m\qquad a=1,\ldots ,5 \,\nonumber\\ K^6{}_6
&\to& K^6{}_6 -H_6\nonumber\\ R &\to &R -{l\over 4} H_6\, ,
\end{eqnarray} which yield the parameter  transformations
\begin{eqnarray}
\label{wef6p} p^{\prime m}&=&p^m+{1\over 2}p^{6}   +{l\over 8}\phi  
\qquad m=1,\dots ,5 \nonumber\\ p^{\prime 6}&=&-{1\over 2}p^6-{3l\over
8}\phi\nonumber \\ \phi^{\prime}&=&{1\over 2} \phi -{2\over l}\,   p^{6}\, .
\end{eqnarray}

\bigskip
\noindent
\underline{\large  $W_7$}

\bigskip
\noindent This Weyl reflexion gives 
\begin{eqnarray}
\label{wef7p} p^{\prime m}&=&p^m \qquad m=1,\dots ,6 \nonumber\\
\phi^{\prime}&=&-\phi\, ,
\end{eqnarray}  which is an $S$-duality like transformation.

\subsection{$G_2^{+++}$}

The Dynkin diagram characterising this non simply laced theory is given   in
Fig.1. The corresponding  maximally oxidised theory is defined in  five
dimensions   with no dilatons and  a two-form field strength.  The action is
given by  
\cite{cremmerjlp99}
\begin{eqnarray}
\label{lag2} S &=&  \int d^5x \, \sqrt{-g^{(5)}}\left(R^{(5)}- {1\over   2.2!
}F_{\mu\nu}F^{\mu\nu} \right)\, + CS .
\end{eqnarray} In addition to  the simple step operators of the gravity   line
one has another step operator
  $R^5$ corresponding to the electric simple root of the two-form. The  
generator
$R^5$ belongs to  a $A_4$-vector $R^a$. The Cartan generators are given   by
\begin{eqnarray}
\label{cartang} H_m&=&(K^m{}_m-K^{m+1}{}_{m+1}) \qquad   m=1,\dots,4\\
\label{cartang5} H_{5}&=&-(K^1{}_1+\ldots +K^{4}{}_{4})+2    K^{5}{}_{5}\, .
\end{eqnarray} The non-trivial Weyl reflexion $W_5$
  corresponding  to the simple roots $\alpha_5$ yields the transformations
\begin{eqnarray}
\label{wegp} p^{\prime m}&=&p^m+p^{5} \qquad m=1,\dots ,4 \nonumber\\ p^{\prime
5}&=&-p^5\, .
\end{eqnarray}

\section{Description of the $KK$ solutions}

\subsection{The $KK$-momentum solution}

We derive  here the group parameters describing the $KK$-momentum solution   in
$D$ dimensions. It is a purely  gravitational uncharged solution.

The $KK$ momentum solution in, say, the  direction $D$ is given by  the
  metric
\begin{equation} ds^2=- H^{-1}\, (dx^1)^2+\sum_{\mu=2}^{D-1}   (dx^\mu)^2+ H
\left[ dx^D +(H^{-1}-1)dx^1 \right]^2,
  \label{kkmetric}
\end{equation} where $H(\{x^\mu\})$ with $\mu=2 \dots D-1$ is a   harmonic
function in $D-2$ dimensions.  In the triangular gauge the vielbein are given by
\begin{eqnarray}
\label{vielkk}e_1{}^{1} &=& H^{-{1\over 2}}\nonumber\\ e_D{}^D &=&   H^{1\over
2}\nonumber\\ e_1{}^D &=&  H^{-{1\over 2}}-H^{1\over 2}\, .
\end{eqnarray} We now use Eq.(\ref{vielbein}) to find the moduli  
$h_a^{~b}$ describing the KK-momentum. For the diagonal components in the  
triangular gauge one has
\begin{equation}
\label{diamod}
  e_i{}^{i}=(e^{-h})_i{}^{i}= e^{-h_i{}^{i}}\, ,
\end{equation} where $i=1$ or $D$. We have thus the following non-zero moduli 
\begin{equation}
\label{cartanmkk} h_1{}^1 = -p^1={1 \over 2} \ln H, \qquad    h_D{}^D =  
-p^D=-{1
\over 2} \ln H,
\end{equation} and 
\begin{equation}
\label{relah} h_1{}^1=-h_D{}^D.
\end{equation} The remaining non-diagonal vielbein gives
\begin{eqnarray}
\label{ndiamod} e_1{}^D &=&(e^{-h})_1{}^D\nonumber\\ &=& -h_1{}^D \left[
1+\sum_{n=2}^{\infty} {1 \over n!} (-1)^{n-1}
\left( \sum_{a=0}^{n-1} (h_1{}^1)^a \, (h_D{}^D)^{n-a-1} \right)
\right] .
\end{eqnarray} Using Eq.(\ref{relah}) in Eq.(\ref{ndiamod}) one gets
\begin{equation}
\label{ndiag} e_1{}^D =-{h_1{}^D \over 2  h_1{}^1} \,  ( e^{h_1{}^1}-  
e^{-h_1{}^1})\, .
\end{equation} Combining Eq.(\ref{ndiag}) with Eqs.(\ref{vielkk}) and
(\ref{cartanmkk}) we  find
\begin{equation}
\label{stepmkk} h_1{}^D= \ln H\, .
\end{equation}  The $KK$-momentum is thus entirely described by the   three
non vanishing group parameters given in Eqs.(\ref{cartanmkk}) and
(\ref{stepmkk}).

\subsection{The $KK$-monopole solution}

We will discuss here the  group parameters describing the $KK$-monopole
solution   in
$D$ dimensions.  The $KK$-monopole solution, in the longitudinal   directions 
$(x^2,
\dots ,x^{D-4})$  and  Taub-NUT  direction $x^D$ is given by  the metric
\begin{eqnarray} ds^2=&-& (dx^1)^2+(dx^2)^2+\dots+ (dx^{D-4})^2 +   H^{-1} (
dx^D + \sum_{i=1}^3 A_i^Ddx^{D-4+i})^2\nonumber\\ &+& H \sum_{i=1}^3
dx^{D-4+i}dx^{D-4+i}\, ,
 \label{kkmonomet}
\end{eqnarray} where $H$ a harmonic function in the 3 transverse   dimensions
 $x^{D-3}, x^{D-2},x^{D-1}$ and
\begin{equation}
\label{potential2}
F_{ij}{}^D\equiv\partial_i A_j^D-\partial_j A_i^D =-\epsilon_{ijk}\, \partial_k
H \qquad i=D-3,D-2,D-1\, .
\end{equation}
 The only non-zero field of this solution which does not
correspond to a generator of   the Cartan subalgebra may be taken to be  
the potential $h_{1\, 2 \dots D-4
\,  D,D}$ of a field dual to $F_{ij}{}^D$. This yields an  `electric' description of the
$KK$-monopole.

To see this let us first consider the  diagonal vielbein.  From
Eq.(\ref{kkmonomet}) the  non-trivial ones  are given by
\begin{eqnarray}
\label{diagveilkm}
  e_D{}^D &=& H^{-{1\over 2}}\nonumber\\ e_i{}^i &=&  H^{1\over 2}
\qquad i=D-3, D-2, D-1 \, ,
\end{eqnarray}
 and the corresponding
moduli 
$p^{(a)}=-h_a{}^a= \ln e_a{}^a$ are
\begin{eqnarray}
\label{modkkmono} p^1&=&0\nonumber\\p^i&=&0 \qquad i=2 \dots D-4\nonumber\\
p^i&=& {1\over 2} \ln H\qquad i=D-3,D-2,D-1\nonumber\\ p^D&=&-{1 \over 2} \ln
H\, .
\end{eqnarray}
The non-diagonal vielbein are
\begin{equation}
\label{nondiag}
e_i{}^D = H^{-{1\over 2}} A_i^D \quad\hbox{or}\quad A_i^D= e_i{}^D
(e^{-1})_D{}^D \qquad i=D-3, D-2, D-1 \, . 
\end{equation}
One  defines a field strength dual to $F_{ij}{}^D$
\begin{equation}
\label{hodgemono}
\sqrt{-g} \widetilde F^{1 \dots \, D-4~ k\,D,D}={1\over 2}
\epsilon^{1 \dots \, D-4~k\,i\,j   D}
F_{ij}{}^D\, .
\end{equation} 
Using the Bianchi identity on the RHS and Eq.(\ref{potential2}) one gets the 
 equation of motion for the $H$-field,
\begin{equation}
\label{motion}
\partial_k \sqrt{-g} \widetilde F^{1 \dots \, D-4~ k\,
D,D}=\sum_i \partial_i \partial_i H =0 \qquad i=D-3, D-2, D-1\, .
\end{equation}
The vielbein matrix and its inverse, evaluated from 
Eqs.(\ref{diagveilkm}) and (\ref{nondiag}) yields $g_{DD}=g^{ii}=H^{-1}$, $g^{ij}=0~~
i\neq j$ and $\sqrt{-g}= H$. Inserting these values in  Eq.(\ref{hodgemono}) and
using Eq.(\ref{potential2}) we see that the quantity
$h_{1\, 2\dots D-4
\,  D,D}$ given by 
\begin{equation}
\label{dualpotential}
h_{1\, 2 \dots D-4
\,  D,D}= \epsilon_{1\, 2 \dots D-4
\,  D}\, (1/H)
\end{equation}
is the potential of the dual field strength
$\widetilde F_{1\, 2 \dots \, D-4~k\, D,D}$.

\section{Step operators and ``dualities'' in $E_8^{+++}$}

In Section 3.3, it has been proven that for all $\G$,  BPS solutions  
transform into each other under  Weyl transformations  of  $\G$. These
transformation   appear in the non-compact dimensions as duality symmetries. 
BPS solutions are  characterised by  only {\em one} field associated to   {\em
one} step operator in addition to the moduli associated to the Cartan  
generators. This is a consequence of   the fact that  all the BPS branes are
always   described as electrically charged  in ${\cal G}^{+++}$. The
transformations of BPS   solutions under Weyl reflexions is thus entirely
determined by the Weyl transformation   of the step operators. In this
Appendix, we illustrate this fact in the familiar case of M-theory, i.e. ${\cal
G}^{+++}=E_8^{+++}$ where the   only non-trivial Weyl reflexion is associated
with the electric root $\alpha_{11}$ (see Fig.1). This transformation has
furthermore an interpretation  in type IIA in terms of  a  double $T$-duality
in the
$(9)$  and $(10)$ directions followed by an exchange of the two directions.  We
shall thus compare the mapping of step operators by Weyl transformations with
these duality transformations.

  We recall that the step operators corresponding to the  simple  roots   of
$E_{8}^{+++}$ are
\begin{equation}
\label{sse8} E_{m}=K^m{}_{m+1}\quad a=1,\ldots ,10 \qquad {\rm and}  
\qquad E_{11}=R^{\,9\,10\,11}\, .
\end{equation} They satisfy  Eq.(\ref{Kcom}) and Eq.(\ref{root}).   Recall that
the representations of $A_{10}$ occuring up to level 3 are  
$(0,0,1,0,0,0,0,0,0,0)$,
$(0,0,0,0,0,1,0,0,0,0)$ and $(1,0,0,0,0,0,0,1,0,0)$. The corresponding fields
and generators are
$h_b{}^a$ and $K^{b}{}_a$ at level 0, $A_{abc}$ and $R^{abc}$ at level   1,
$A_{abcdef}$ and $R^{abcdef}$ at level 2 and finally
$h_{a_1 \dots a_8,b}$ and $R^{a_1 \dots a_8,b}$ at level 3 which are   totally
antisymmetric in the
$a$ indices and the antisymmetrised part in all its indices vanishes  
\cite{west01, damourhn02}. In the normalisation Eq.(\ref{normalisation}) we
have the   following relation between the operators of level $1$ and level $2$
\begin{equation}
\label{relcom} [R^{abc},R^{def}]=R^{abcdef}\,  .
\end{equation}

The action of the electric Weyl reflexion $W_{11}$ on the Cartan   subalgebra is
given in Eq.(\ref{alpha11}). On   the
 positive generators belonging to $GL(D)$, one gets,  using Eq.(\ref{Kcom}) 
and Eq.(\ref{root}), 
\begin{eqnarray}
\label{kw11a} K^{\prime i}{}_{j}&=&  K^i{}_{j} \qquad i<j \leq 8  
\nonumber\\ K^{\prime 9}{}_{10}&=&K^9{}_{10}  \qquad\qquad K^{\prime
10}{}_{11}=K^{10}{}_{11}
\nonumber\\ K^{\prime i}{}_{9}&=&R^{\,  i\,10\,11}  ~~\qquad\quad   K^{\prime
i}{}_{10} =R^{\, i\,11\,9}
\quad i=1\dots 8\\ \label{kw11aa}
\label{kw11b}K^{\prime i}{}_{11}&=&R^{\, i\, 9\, 10} \qquad i=1 \dots   8\, .
\end{eqnarray} These transformations are in agreement with the interpretation
of $W_{11}$ as a double $T$-duality plus
echange in the
$(9)$ and $(10)$ directions. Indeed in type $IIA$ a non-zero
$h_1{}^9$ or $h_1{}^{10}$ corresponds to a $KK$-momentum in the $(9)$ or $(10)$
direction (see Appendix B).  Performing  the double duality plus the exchange
give a $1F$ in the $(10)$  or $( 9)$ direction. Uplifting back to M-theory this
gives a $M2$-brane in  $(10\, 11)$  or   in the directions $(9\, 11)$ described
by a non-vanishing
$A_{\, 1\, 10\,11}$ or $A_{\, 1\, 9\, 11}$  which is consistent with  
Eq.(\ref{kw11a}) with
$i=1$. Similarly a non-zero $h_1{}^{11}$ describes in type $IIA$ to a  
$D0$ brane which gives after the duality and uplifting a $M2$-brane in the
$(9\,   10)$ directions, in agreement with  Eq.(\ref{kw11aa}).

The transformations of the remaining $R^{abc}$ under $W_{11}$ give
\begin{eqnarray}  R^{\prime \,9\,10\,11}&=&R_{\,9\,10\,11}\nonumber\\
\label{r3w11a} R^{\prime ija}&=& - R^{ija}\qquad i,j \leq 8, \quad a  
 >8\\
\label{r3w11b} R^{\prime ijk}&=&R^{ijk\,9\,10\,11}  \qquad i,j,k \leq   8\, .
\end{eqnarray} The first relation simply states that under $W_{11}$,  
$\alpha_{11}$ is the only positive root mapped to a negative one. The second
and third relations  are consistent with the $T$-duality interpretation on the  
branes. Consider first Eq.(\ref{r3w11a}) with a time index.  The non zero
$A_{1ia}$ describes an $M2$-brane in the $(ia)$ directions. There are two cases
to distinguish:
$a=11$ and $a=9$ or $10$. If $a=11$ it corresponds in the $IIA$   language to a
$1F$ in the $(i)$ direction which is indeed invariant under the double
$T$-duality plus exchange. In the case $a=9$ we have a  
$D2$-brane\footnote{The same  argument can be repeated for $a=10$.} in the
$(i9)$ directions which upon the double $T$-duality in the $(9\,10)$  
directions gives a $D2$ in
$(i10)$, and the exchange of the $(9)$ and $(10)$ directions gives back the
original $D2$. The uplifted $M2$ is thus invariant and   Eq.(\ref{r3w11a}) is
consistent with the duality picture on the $M2$ brane. Consider now
Eq.(\ref{r3w11b}) with atime index. The
$M2$   brane lying in the $(ij)$ directions is    described by  a non zero
$A_{1ij}$. This  
$M2$ yields a $D2$ in type
$IIA$ that under the double $T$-duality plus exchange is mapped onto a
$D4$ brane in the direction $(ij\,9\,10)$. When uplifted to $11$
dimensions it yields an $M5$ in the $(ij\,9\,10\,11)$ directions described   by
a non-zero  $A_{1ij\,9\,10\,11}$,  in agreement with  
Eq.(\ref{r3w11b}).

We finally analyse the transformations of the remaining $R^{abcdef}$   under
$W_{11}$. We get
\begin{eqnarray}
\label{r6w11a} R^{\prime ijklab}&= & R^{ijklab} \qquad i,j \leq 8,  
\quad a >8\\
\label{r6w11b} R^{\prime ijklma}&=&R^{ijklmbca,a}  \qquad i,j,k \leq 8
\quad a \neq b \neq c >8.
\end{eqnarray} Note first that we did not list the transformation of  
$R^{a_1 \dots a_6}$ with all the six indices  $a_i \leq 8$. These yield 
operators of level 4. The associated parameters do not   have an obvious 
interpretation in terms  of eleven-dimensional   supergravity fields and 
their interpretation in terms of branes has not yet been clarified. This is not
too surprising because it corresponds in the usual   U-duality discussion of
M-theory to a case where the transverse non-compact space  is of   dimension
two and where the significance of the U-duality orbits is unclear
\cite{obersp98}.

Consider first Eq.(\ref{r6w11a}).  This   equation
implies that an $M5$ with two longitudinal directions     
$(9), (10)$ or
$(11)$is invariant under the Weyl transformation. There are two cases.  First,
none of the two directions is $(11)$. In type
$IIA$ one   has then a
$NS5$ longitudinal in  $(9\, 10)$, which is indeed separately invariant under
the double
$T$-duality and under the exchange.  Second, one of the two directions is
$(11)$. In this case upon dimensional reduction one gets   a $D4$ along one of
the two $(9),(10)$ directions, which is also invariant under the conjugate  
action of the double $T$-duality and of the exchange of the $(9)$ and $(10)$
directions.

We now consider  Eq.(\ref{r6w11b}). It is  
relevant for an $M5$ with only one longitudinal direction $(a)$ 
along  
$(9), (10)$ or  $(11)$.  The case $a=11$  corresponds  in type $IIA$ 
 to a $D4$. Performing  the double $T$-duality plus exchange one   gets a
$D6$ which is uplifted  in M-theory to a $KK6$ monopole  with $a=11$ as the  
Taub-NUT direction. The case $a=9$ (the discussion is similar
for  
$a=10$)  corresponds in type $IIA$ to a $NS5$ brane with $a=9$ as one of its 
longitudinal directions The $T$-duality in the longitudinal
direction
$a=9$ leaves the $NS5$ invariant but the $T$-duality in the transverse
$(10)$ direction maps the $NS5$ onto a $KK5$ monopole with  (10) as the
Taub-NUT direction. Finally, the exchange
$( 9) \leftrightarrow (10)$  selects $a=9$ as the Taub-NUT direction 
of the $KK5$. Thus  the interpretation of   the Weyl reflexion
$W_{11}$ as a duality is consistent with the  electric description of the
$KK6$-monopole of M-theory in terms of  only {\it   one}
non-vanishing field  of level 3, namely the field $h_{t\,y_1 \dots y_6\,b,b}$
associated to the step operator
$R^{t\,y_1 \dots y_6\,b,b}$ with  $y_i$ as the longitudinal   coordinates
and $b$ as the Taub-NUT direction. The KK-monopole being a purely gravitational
solution, it does  not   depend on the possible presence of form field strengths
and  may, for all $\G$, be similarly characterised  by a field
$h_{t\,y_1 \dots y_{D-5}\,b,b}$ with  $y_i$ as the longitudinal  
coordinates and $b$ as the Taub-NUT direction.

\end{document}